\documentstyle[aps]{revtex}

\input{epsf.tex}

\begin{document}
\draft
\title{Effect of Dynamical $SU(2)$ Gluons to the Gap Equation of Nambu--Jona-Lasinio Model in Constant Background Magnetic Field}
\author{
Masaru Ishi-i\thanks{isii1scp@mbox.nc.kyushu-u.ac.jp}, 
Taro Kashiwa\thanks{taro1scp@mbox.nc.kyushu-u.ac.jp}
}
\address{Department of Physics, Kyushu University, Fukuoka, 812-8581, Japan}
\author{Naoki Tanimura\thanks{ntanimur@star.fuji-ric.co.jp}}
\address{Fuji Research Institute Corporation, Chiyodaku, Tokyo, 101-8443, Japan}
\date{\today}
\maketitle
\begin{abstract}
In order to estimate the effect of dynamical gluons to chiral
condensate, the gap equation of $SU(2)$ gauged Nambu--Jona-Lasinio model, under a constant background magnetic field, is investigated up to the two-loop order in $2+1$ and $3+1$ dimensions. We set up a general formulation allowing both cases of electric as well as magnetic background field. We rely on the proper time method to maintain gauge invariance. In $3+1$ dimensions chiral symmetry breaking ($\chi$SB) is enhanced by gluons even in zero background magnetic field and becomes much striking as the background field grows larger. In $2+1$ dimensions gluons also enhance $\chi$SB but whose dependence on the background field is not simple: dynamical mass is not a monotone function of background field for a fixed four-fermi coupling.
\end{abstract}

\section{Introduction}
The Nambu--Jona-Lasinio (NJL) model \cite{NJL}, the model of four fermion interaction, has been discussed as a possible realistic mechanism of chiral symmetry breaking ($\chi$SB)\cite{TC,TQC}. Although the interaction is nonrenormalizable in more than $1+1$ spacetime dimensions, it is regarded as a low energy effective theory of elementary fermions after integrations of Higgs and/or gauge fields, and has been used as a phenomenological model to describe hadronic spectra, decays, and scattering
\cite{DWER}. (While these properties of hadrons should be derived from quantum chromodynamics, analytical methods on hand are so limited that dynamics in the low-energy region could not be easily explored.)

In order to grasp qualitative behavior of a system, it is sometimes useful to examine response to external fields or sources. Many such attempts have been made so far, for example, the $O(3)$ gauge-Higgs model in a magnetic field source\cite{Linde}, fermionic models minimally coupled to strong electromagnetic fields\cite{ST1}, the NJL model minimally coupled to a constant magnetic field and curved spacetime\cite{NJLGR} and its extension to supersymmetric NJL model\cite{SUSYNJLGR}, the instanton motivated four-point interaction model of fermion at finite density\cite{Instanton}, and so forth.

Among these, the most interesting outcome is found in the study of the NJL model minimally coupled to constant external magnetic fields\cite{KL,ST2,IKT}; where mass generation occurs even at the weakest attractive interaction between fermions in terms of so called the ``dimensional reduction(DR)''\cite{GMS1}. The case is also investigated by the present authors to find that the origin of DR is the infrared divergences followed from the fermion loop integral under the influence of the external magnetic fields\cite{IKT}. The phenomenon is now understood universal provided interactions are shortrange\cite{SSW}.

Inclusion of dynamical gauge fields is made for QED, in terms of the Schwinger--Dyson equation\cite{GMS2} or renormalization group(RG)\cite{HKS}. The results show that the dynamical symmetry breaking always occur with the aid of an external magnetic field.

The motive for this work is traced to that of Gusynin, Hong, and Shovkovy\cite{GHS}; where $2+1$-dimensional $SU(2)$ gauge theory is investigated in a constant magnetic field by using a constant gauge potential with translational invariance\cite{LW}, to find that magnetic catalysis of $\chi$SB does not occur. The study in 3+1 dimensions is made in the reference \cite{BGO}, where $SU(N)$ gauged NJL model is handled in the case of weak as well as constant magnetic field: they incorporate dynamical effect of gluons in RG to reach the conclusion that existence of the external field does not change the condition of $\chi$SB. Each result indicates, contrary to our expectation that gluons always trigger $\chi$SB (since they do in the low energy region), that dynamical gluons in external magnetic field do not play a major role to $\chi$SB. This curious situation is our starting point. There needs a more detailed analysis.

In this article we study $\chi$SB under a constant background magnetic field in $SU(2)$ gauged NJL model, paying attention to (i) the direct effect of dynamical gluons to the gap equation, not in terms of RG\cite{GHS,BGO},  and (ii) the effect of gauge choice to the results: since in nonabelian case the situation, $B_z^3=B \ (\mbox{constant}\geq 0)  \ \ ; \  \mbox{others}=0$, the choice\cite{GHS}
\begin{equation}
A_x^1 = A_y^2 = \sqrt{B} \ ; \quad  \mbox{others} = 0 \ ,
\label{nonCovariantChoice}
\end{equation}
cannot connect with \cite{BGO},
\begin{equation}
A_x^3 = -\frac{  B }{  2  }y \ , \quad A_y^3 = \frac{  B }{  2  }x \ ,  \quad  
\mbox{others} =0 \ :
\label{CovariantChoice}
\end{equation}
since the remnant gauge transformation which leaves $B_{z}^{3}$ invariant, that is, any gauge transformations with respect to the third axis cannot bring the 
gauge (\ref{nonCovariantChoice}) to the gauge (\ref{CovariantChoice}) or vise versa. To work with the Green's function in momentum space it is convenient to adopt eq.(\ref{nonCovariantChoice}), but this is not a solution of the nonabelian Maxwell equation in the vacuum,
\begin{equation}
(D_\mu F_{\mu \nu})^a \equiv \partial_\mu F_{\mu \nu}^a +
\epsilon^{abc}A_{\mu}^{b}F_{\mu \nu}^c = 0 \ . 
\label{MaxwellEquation}
\end{equation}
In what follows we rely upon the WKB semi-classical approximation where the classical solution plays a fundamental role. Therefore we should work with the choice eq.(\ref{CovariantChoice}), satisfying eq.(\ref{MaxwellEquation}). Moreover the choice (\ref{CovariantChoice}) fulfills a covariantly constant condition,
\begin{equation}
(D_\rho F_{\mu \nu})^a \equiv \partial_\rho F_{\mu \nu}^a +
\epsilon^{abc}A_{\rho}^{b}F_{\mu \nu}^c = 0 \ ,
 \label{CovariantConstant}
\end{equation}
enabling us to use the Fock--Schwinger proper time method\cite{FS} that was originally developed to handle abelian background fields and that gives us a gauge independent result.

For calculations of the functional determinant, a regularization for ultraviolet divergence is necessary. We introduce a proper time cutoff, which preserves gauge symmetry. In addition, there needs another cutoff: in order to estimate the contribution from dynamical gluons, two-loop order of the effective potential must be taken into account. The gluon propagator under the magnetic background suffers from the famous instability, due to tachyonic modes of the spin $1$ propagator under the background magnetic fields\cite{NO}. To circumvent this instability we introduce a gluon mass whose square is larger than the magnitude of a magnetic field.

The paper is organized as follows: in Sec.~\ref{S2} a general formulation is presented, where we assume generic backgrounds satisfying the covariantly constant condition eq.(\ref{CovariantConstant}). In the next sections, Sec.~\ref{S3} and Sec.~\ref{S4}, the gap equations in $2+1$ and $3+1$ dimensions are shown. The last section Sec.~\ref{S5} is devoted to discussion. In Appendix~\ref{Appa} calculations of the kernel is presented and in Appendix~\ref{Appb} the gluon propagator is represented by means of the proper time method. Then in Appendix~\ref{Appc} we make an explicit proof that our classical solution does satisfy the covariantly constant condition. And finally in Appendix~\ref{Appd}, in order to ensure the gauge independence of our calculations, that is, the correctness of those, we study the Ward--Takahashi relation of vacuum polarization function.

\section{Formulation}\label{S2}
In this section, we derive the effective potential of $SU(2)$ gauged NJL model. The Lagrangian of the system in the Euclidean formulation is given as
\begin{eqnarray}
{\cal L}= -{1\over 4e^2}F_{\mu\nu}^a F_{\mu\nu}^a
-\overline\psi\left\{\gamma_{\mu}
\left(\partial_{\mu}-iA_{\mu}\right)\right\}\psi
+{g^{2}\over2}\left\{\begin{array}{ll}
\left[\bigl(\overline\psi\psi\bigr)^{2}
+\bigl(\overline\psi i\gamma_{5}\psi\bigr)^{2}\right]\ ,&D=4\ ,\\
\noalign{\vspace{1ex}}
\left[\bigl(\overline\psi\psi\bigr)^{2}
+\bigl(\overline\psi i\gamma_{4}\psi\bigr)^{2}
+\bigl(\overline\psi i\gamma_{5}\psi\bigr)^{2}\right]\ ,&D=3\ ,
\end{array}\right.
\label{GNJL}
\end{eqnarray}
where $A_{\mu}\equiv A^{a}_{\mu}T_a$ with $T_a$'s $(a=1,2,3)$ being the  $SU(2)$ generators given by $T_a \equiv \sigma_{a}/ 2$, where $\sigma_{a}$'s are the Pauli matrices. For the $2+1$-dimensional case, a spinorial representation of the Lorentz group is given by two-component spinors, so that corresponding gamma matrices are $2\times2$. There is no chiral symmetry. In order to be able to discuss chiral symmetry, we introduce an addtional flavor such that
\begin{equation}
\psi=\left(\begin{array}{c}\psi_{1}\\\psi_{2}\end{array}\right)\ ,\quad
\overline\psi\equiv\psi^{\dag}\gamma_{3}
\equiv\left(\begin{array}{cc}\overline\psi_{1},
&-\overline\psi_{2}\end{array}\right)
\equiv\left(\begin{array}{cc}\psi_{1}^{\dag}\sigma_{3},
&-\psi_{2}^{\dag}\sigma_{3}
\end{array}\right)\ ,
\end{equation}
with the $4\times4$ gamma matrices
\begin{equation}
\gamma_{\mu}=\left(\begin{array}{cc}\sigma_{\mu}&0\\
0&-\sigma_{\mu}\end{array}\right)\,;\mu=1\sim3\,,\quad
\gamma_{4}=\left(\begin{array}{cc}0&{\bf1}\\{\bf1}&0\end{array}\right)\,,
\quad\gamma_{5}=\gamma_{1}\gamma_{2}\gamma_{3}\gamma_{4}
=\left(\begin{array}{cc}0&i{\bf1}\\-i{\bf1}&0\end{array}\right)\ .
\label{gamma3d}
\end{equation}
Chiral symmetry is realized as
\begin{equation}
\psi\longrightarrow{e}^{i\alpha\gamma_{4}}\psi\,,\quad
\psi\longrightarrow{e}^{i\beta\gamma_{5}}\psi\ ,
\end{equation}
yielding a global $U(2)$ symmetry which is broken by a mass term into $U(1)\times U(1)$.

The partition function of the model is read as
\begin{eqnarray}
Z&\equiv&\int\!\!{D}[\mbox{gauge}]
{ D}[\psi]{ D}[\overline\psi]
\exp\left[\!\int\!\!d^{D}x\,  {\cal L}  \right]= \int\!\!{
D}[\mbox{gauge}]{ D}[\sigma]{ D}[{\bbox\pi}]
{D}[\psi] {D}[\overline\psi]
\nonumber \\
& & \times \exp\left[-\!\int\!\!d^{D}x
\left\{{1\over 4e^2}F_{\mu\nu}^a F_{\mu\nu}^a
+{1\over2g^{2}}\bigl(\sigma^{2}+{\bbox\pi}^{2}\bigr)
+\overline\psi\left\{\gamma_{\mu}\bigl(\partial_{\mu}-iA_{\mu}\bigr)
+\bigl(\sigma+i{\bbox{\pi\cdot{\mit\Gamma}}}\bigr)\right\}\psi
\right\}\right] \ ,
\end{eqnarray}
where the auxiliary fields, $\sigma$ and ${\bbox\pi}$, have been introduced to erase the four-fermi interactions,
\begin{equation}
{\bbox\pi}\cdot{\bbox\Gamma}=\left\{
\begin{array}{ll}\pi\gamma_{5}\,&\mbox{for }D=4\,,\\
\pi_{1}\gamma_{4}+\pi_{2}\gamma_{5}\,&\mbox{for }D=3\ ,
\end{array}\right.
\end{equation}
and a measure of gauge fields, ${D}[\mbox{gauge}]$, is specified after the following procedures: (i)~first, integrate with respect to fermions to give
\begin{eqnarray}
Z=\int\!\!{D}[\mbox{gauge}]{D}[\sigma]{D}[{\bbox\pi}]
\exp\left[-\!\int\!\!d^{D}x
\left\{ {1\over 4e^2}F_{\mu\nu}^a F_{\mu\nu}^a
+{1\over2g^{2}}\bigl(\sigma^{2}+{\bbox\pi}^{2}\bigr) \right\}
+{\rm Ln}{\rm{Det}}\Bigl[ \gamma_{\mu}\bigl(\partial_{\mu}-iA_{\mu}\bigr)
+\bigl(\sigma+i{\bbox{\pi\cdot{\mit\Gamma}}}\bigr) \Bigr] \right] \ .
\end{eqnarray}
(Here and hereafter Ln, Det, and Tr designate the functional logarithm, determinant, and trace, respectively.) (ii)~Second, set up an ansatz; $\sigma (x)=m (:\mbox{constant}), {\bbox\pi}=0$, under which the equation of gauge fields reads,
\begin{eqnarray}
{1\over e^2}\Big(\partial_\mu {\cal F}_{\mu \nu}^a + \epsilon^{abc}
{\cal A}_{\mu}^b {\cal F}_{\mu \nu}^c\Big) 
\equiv{1\over e^2} ({\cal D}_\mu {\cal F}_{\mu\nu})^a
=-{\rm tr} \Big[\big(\gamma_{\mu}(\partial_{\mu}
-i{ \cal A}_{\mu})+m \big)^{-1}\big(-i\gamma_{\nu}T_a \big)  \Big] \ .
\label{EQUATION}
\end{eqnarray}
If we take ${\cal F}_{\mu\nu}^1={\cal F}_{\mu\nu}^2=0 \ ; {\cal F}_{\mu\nu}^3 = \mbox{constant}$ , then
\begin{equation}
{\cal A}_{\mu}^i = 0 \ ; (i=1,2) \ : \qquad {\cal A}_{\mu}^3=-{1\over2}{\cal F}_{\mu\nu}^3 x_{\nu} \ ;  
\label{ClassicalSolution}
\end{equation}
the equation (\ref{EQUATION}) is fulfilled. (It is easy to convince that the left-hand-side vanishes but the proof of the right-hand-side is rather lengthy, so it is relegated to Appendix \ref{Appc}.) Therefore our classical solution 
${\cal A}_{\mu}^a $(\ref{ClassicalSolution}) obeys the nonabelian Maxwell equation (\ref{EQUATION}). We do not take the higher powers of $\delta \sigma$ and $\delta \pi$, 
\begin{eqnarray}
\sigma (x) = m + \delta \sigma (x) \ ,  \qquad  \pi (x) = 0 + \delta \pi(x)  \ ;
\end{eqnarray}
but remain the lowest part. In this section we do not restrict ourselves in the pure magnetic case but in generic cases where both electric and magnetic backgrounds coexist.(iii)~Third, expand the gauge fields around ${\cal A}_{\mu}^a$
\begin{equation}
A_{\mu}^a={\cal A}_{\mu}^a+Q_{\mu}^a \ ,
\end{equation}
with $Q^a_{\mu}$ being designated as quantum fields. Here, with the aid of the Faddeev-Popov trick, the measure is defined as
\begin{equation}
{D}[\mbox{gauge}] \equiv { D}\left[Q_{\mu}^a \right]
\biggl\vert {\rm{Det}} {\delta G^a \over \delta \theta^b}\biggr\vert
\exp\left[-{1\over 2 e^2}
\int d^{D}x\bigl(G^a(x)\bigr)^2 \right ] \ ,
\end{equation}
with
\begin{equation}
G^a(x)\equiv({\cal D}_{\mu}Q_{\mu})^a=\partial_{\mu}Q_{\mu}^a+ \epsilon^{abc}
{\cal A}_{\mu}^b Q_{\mu}^c \ .
\end{equation}
The gauge transformation now reads,
\begin{equation}
Q_{\mu}^a\mapsto Q_{\mu}^a+(D_{\mu}\theta)^a \ ,
\end{equation}
with $D_{\mu}^{ab}\equiv \delta^{ab}\partial_{\mu}
+\epsilon^{acb}({\cal A}_{\mu}^{c}+Q_{\mu}^{c})$,
so that the Faddeev-Popov determinant is given by
\begin{equation}
{\rm{Det}} {\delta G^a \over \delta \theta^b} = {\rm{Det}}(D_\mu
{\cal D}_\mu)^{a b} \ .
\end{equation}
(In what follows, however, it is not necessary to worry about the F-P terms; since they are irrelevant to the gap equation.)

The partition function is given, up to $O(Q^2)$, by
\begin{eqnarray}
Z[A]&=&\exp\left[-\int d^{D}x {m^2 \over 2g^2}
+ {\rm Ln}{\rm{Det}}\left[\gamma_\mu(\partial_\mu-i{\cal A}_\mu^3 T_3)
+m\right]\right]
\times
\exp\left[-\!\int\!\!d^{D}x {1\over 4e^2}
{\cal F}_{\mu\nu}^3 {\cal F}_{\mu\nu}^3 \right]
\nonumber  \\
&& \hspace{-4ex} \times\int { D}\left[Q_{\mu}^a
\right]\exp\left[-{1\over2e^2}\int d^{D}xd^{D}y \ Q_\mu^a
\left(\Delta^{-1}_{\mu\nu}+e^2\Pi_{\mu\nu}+e^2(\mbox{F-P term})\right)^{a b}
Q_\nu^b \right]  \ ,
\label{PartitionFunction1}
\end{eqnarray}
where $\Delta^{-1}$ is the inverse of the gluon propagator,
\begin{equation}
(\Delta^{-1})^{ab}_{\mu\nu}\equiv-\delta_{\mu\nu}({\cal D}^2)^{ab}
+2\epsilon^{ab3}{\cal F}^3_{\mu\nu} \ , \label{GluonPropagator}
\end{equation}
under the gauge (\ref{ClassicalSolution}). In eq.(\ref{GluonPropagator}), the symmetric matrix $\epsilon^{ab3}{\cal F}^3_{\mu\nu}$ has negative eigenvalues after the diagonalization\cite{NO}, thus there are tachyonic singularities. This is due to a large magnetic moment of spin $1$ particle. (Recall that we are in the Euclidean world so that all gauge fields are considered as magnetic one.) In view of eq.(\ref{GluonPropagator}), these tachyonic singularities become harmless if we introduce a gluon mass $M_g$ that should obey
\begin{eqnarray}
 (M_g)^2 >  |{\cal F}_{\mu \nu}^3|   \ . 
 \label{GluonMassCondition}
\end{eqnarray}

The term in eq.(\ref{PartitionFunction1}),
\begin{equation}
\Pi_{\mu \nu}^{a b} \equiv  - \frac{ \delta^2 }{ \delta Q_\mu^a
\delta Q_\nu^b    }{\rm Ln}{\rm{Det}}\Big[ \gamma_{\mu}
\bigl(\partial_{\mu} - i{\cal A}_\mu^3 T_3 -iQ_\mu \bigr) + m  \Big]
\Bigg\vert_{Q=0}\hspace{-3ex}  =  - \frac{ \delta^2 }{ \delta Q_\mu^a
\delta Q_\nu^b    }{\rm Tr} {\rm Ln} \Big[ \gamma_{\mu}
\bigl(\partial_{\mu} - i{\cal A}_\mu^3 T_3 -iQ_\mu \bigr) + m  \Big]
\Bigg\vert_{Q=0} ,
\label{VacPolTensor}
\end{equation}
is the vacuum polarization tensor. We omit the isospin index $3$ of ${\cal A}^3_\mu$ as well as ${\cal F}^3_{\mu \nu}$ from now on.

Integrating with respect to the quantum field $Q_{\mu}$, we obtain
\begin{eqnarray}
Z=\exp\left[-VT{\mbox{\Large$v$}}^{(D)} \right] \times (m-\mbox{independent
terms}) \ ,
\end{eqnarray}
where
\begin{eqnarray}
&&{\mbox{\Large$v$}}^{(D)}\equiv{m^2 \over 2g^2}+{\mbox{\Large$v$}}_1^{(D)}
+{\mbox{\Large$v$}}_2^{(D)} \ ,  \label{TotalEffectivePot}
\end{eqnarray}
with
\begin{eqnarray}
&&{\mbox{\Large$v$}}_1^{(D)}\equiv-{1\over VT}{\rm Tr} {\rm Ln} 
\Bigl[\gamma_{\mu}
\bigl(\partial_{\mu}-i{\cal A}_\mu T_3\bigr)+m\Bigr]  \ ,
\label{1loopEffectivePot}\\
&&{\mbox{\Large$v$}}_2^{(D)}
\equiv  {1\over VT}{e^2\over 2}{\rm{Tr}}({\bbox \Pi}{\bbox\Delta})
= {1\over VT}{e^2\over 2} \int d^{D}x
d^{D}y \ \Pi_{\mu\nu}^{ab}(x,y)\Delta_{\nu\mu}^{ba}(y,x)  \ , 
\label{2loopEffectivePot}
\end{eqnarray}
being the one-loop and the two-loop effective potential respectively. Here $V$ is the $(D-1)$-dimensional volume of the system, $T$ is the Euclidean time interval. The stationary condition for the effective potential,
\begin{equation}
{\partial \mbox{\Large$v$}^{(D)} \over \partial m}=0 \ ,
\end{equation}
gives a gap equation,
\begin{eqnarray}
-{(4\pi)^{D/2} \over 4g^2\Lambda^{D-2}}=f_1^{(D)}(x)+f_2^{(D)}(x)
\quad \ ; \qquad x\equiv{m^2\over {\Lambda}^2} \ , \qquad  \left( 0 \leq x \leq 1 \right)  \ ; \label{TheGap}
\end{eqnarray}
where $\Lambda$ is the ultraviolet cut-off, and
\begin{eqnarray}
f_i^{(D)}(x)&\equiv&{(4\pi)^{D/2} \over 2\Lambda^{D-2}}
\frac{\partial {\mbox{\Large$v$}}_i^{(D)}}{\partial m^{2}} \ ; \qquad i=1,2 \ .
\label{fvRelation}
\end{eqnarray}

The remaining task is therefore the calculation of the effective potential eqs.(\ref{1loopEffectivePot}) and (\ref{2loopEffectivePot}). Let us start with the one-loop part. We utilize the proper time method\cite{FS}:
\begin{eqnarray}
{\mbox{\Large$v$}}_1^{(D)}&=&-{1\over VT}{\rm{Tr}}{\rm{Ln}}\Bigl[\gamma_{\mu}
\bigl(\partial_{\mu}-i{\cal A}_\mu T_3\bigr)+m\Bigr] = -{1\over
2VT}{\rm{Tr}}{\rm{Ln}}
\Bigl[-\bigl(\partial_{\mu}-i{\cal A}_\mu T_3\bigr)^{2}
-{1 \over 2}\sigma_{\mu\nu}{\cal F}_{\mu\nu} T_3+m^2\Bigr] \nonumber  \\
&=& \frac{  1 }{ 2 VT   } \int^{\infty}_{1/\Lambda^{2}}\!d\tau \
\tau^{-1} {e}^{- \tau m^2}
\! {\rm tr}\!  \int \! d^Dx \langle x  | {e}^{- \tau H_0} | x
\rangle  \ ,
\label{1LoopEffectivePot}
\end{eqnarray}
where the ultraviolet cutoff $\Lambda$ has been introduced, tr is taken only for the spinor and the isospin indices, and
\begin{equation}
H_0 \equiv {\Pi_{\mu}}^2 -{1\over 2}\sigma_{\mu\nu}{\cal F}_{\mu\nu}T_3 \ ,
\label{H_0definition}
\end{equation}
with
\begin{equation}
\Pi_{\mu}\equiv \hat{p}_\mu - {\cal A}_\mu(\hat{x})T_3 \ ,
\label{OperatorPi}
\end{equation}
$[\hat{x}_\mu, \hat{p}_\nu] = i\delta_{\mu \nu}$. Write the kernel as
\begin{equation}
K(x,y; \tau) \equiv\langle x  | {e}^{- \tau H_0} | y \rangle \ ,
\label{kernel}
\end{equation}
to find
\begin{eqnarray}
K(x,y; \tau)   & \!  =  \!  & \frac{  1 }{ (4 \pi \tau)^{D/2}   }
\left[{\rm det} \left(\frac{ \sin \tau {\cal F} /2 }{  \tau {\cal F}/2  }
\right)_{\mu \nu}
\right]^{-1/2}
\big( K_0(\tau) {\bf I}  +    K_3(\tau) T_3 \big) \exp
\left[ i T_3 {\cal C} \right]   \nonumber  \\
&\times&  \exp \big[
        -\frac{  1 }{ 4   } (x-y)_\mu \left(\frac{{\cal F} }{ 2   } \cot
\frac{ \tau{\cal F} }{ 2 } \right)_{\mu \nu}  (x-y)_\nu \big]  \ ,
\label{kenel'sExpression}
\end{eqnarray}
whose derivation is given in Appendix~\ref{Appa}. In eq.(\ref{kenel'sExpression}), quantities are defined such that
\begin{eqnarray}
& & {\cal C}\equiv-\frac{1}{2}{\cal F}_{\mu\nu}x_{\mu}y_{\nu}  \ ,
\label{A-no-Definition} \\
\quad  K_0(\tau)    & \! \equiv  \!  &
\cosh \frac{ \tau F_+  }{  2  } \cosh \frac{ \tau F_-  }{  2  }
        - \gamma_5  \sinh \frac{ \tau F_+  }{  2  } \sinh \frac{ \tau F_-  }
{  2  }  \ ,
\\
K_3(\tau)& \! \equiv   \!  & \sigma_{\mu \nu}
\Big(N_{\mu \nu}^{+} \sinh \frac{ \tau F_+  }{  2  } \cosh \frac{ \tau F_-  }
{  2  }  + N_{\mu \nu}^{-} \cosh \frac{ \tau F_+  }{  2  }
\sinh \frac{ \tau F_-  }{  2  }    \Big)  \ ;
\end{eqnarray}
where
\begin{eqnarray}
& \!  \!  & \left\{ \begin{array}{l}
F_+  \equiv \sqrt{B^{2}+{\bbox E}^{2}}  \\
\noalign{\vspace{1ex} }
F_-  =  0
\end{array} \right.  \ ; \hspace{8ex} B \equiv {\cal F}_{12} \ ; {\bbox
E}\equiv ({\cal F}_{13}, {\cal F}_{23}) \ ; \quad   \mbox{for $D=3$} \ ;
    \label{Fs1}  \\
    & \!  \!  &  \left\{ \begin{array}{l}
\displaystyle{ {F}_{+} \equiv  \frac{ \left\{ \vert{\bbox{B+E}}\vert +
\vert{\bbox{B-E}}\vert\right\}  }{  2  } } \ ; \ {\bbox E} \equiv ({\cal
F}_{14},{\cal F}_{24},{\cal F}_{34}) \ ; \\
\noalign{\vspace{1ex} }
\displaystyle{ F_- \equiv \frac{  \left\{\vert{\bbox{B+E}}\vert -
\vert{\bbox{B-E}}\vert\right\} }{  2  } } \ ;  \ {\bbox B} \equiv ({\cal
F}_{23},{\cal F}_{31},{\cal F}_{12}) \ ;
\end{array} \right.  \ ; \quad  \mbox{for $D=4$} \ ;  \label{Fs2}
\end{eqnarray}
and
\begin{eqnarray}
      N_{\mu\nu}^{+} \equiv \frac{ {\cal F}_{\mu \nu}  }{  F_+  } \equiv
N_{\mu \nu}  \quad  & ;
&  \quad
N^{-}_{\mu\nu} = 0  \    ;    \hspace{5ex}  \mbox{for $D=3$} \ ;
\label{Nin3-dim} \\
      N_{\mu\nu}^{+} \equiv \frac{ {\cal F}_{\mu \nu} F_+ - \tilde{\cal F}_{\mu
\nu} F_- }
{  F_+^2 - F_-^2  } \quad  & ; &  \quad N_{\mu\nu}^{-}\equiv \frac{
\tilde{\cal F}_{\mu \nu}
F_+ - {\cal F}_{\mu \nu} F_- }{  F_+^2 - F_-^2  }  \  ;   \quad
\mbox{for $D=4$} \ ; \label{Nin4-dim}
\end{eqnarray}
with $\tilde{\cal F}_{\mu \nu} \equiv \epsilon_{\mu \nu \lambda \rho}{\cal F}_{\lambda \rho}/2$ being the dual of ${\cal F}_{\mu \nu}$.

Therefore
\begin{equation}
{\mbox{\Large$v$}}_1^{(D)}= \frac{  1 }{ 2 VT   }
\int^{\infty}_{1/\Lambda^{2}}\!d\tau \   \tau^{-1} {e}^{- \tau m^2}
\! {\rm tr}\!  \int \! d^Dx K(x,x; \tau) =
{4\over(4\pi)^{D/2}}
\int^{\infty}_{1/\Lambda^{2}}\!d \tau\,\tau^{-D/2-1}
{e}^{-\tau m^{2}}G_{D}(\tau F) \ ,
\label{Ir}
\end{equation}
where
\begin{equation}
G_{D}(\tau F) \equiv
\left\{
\begin{array}{lcl}
\displaystyle{
\frac{\tau F_+}{2}
\coth \! \left( \frac{\tau F_+ }{ 2 } \! \right)}
& {\mathrm{for}} &  D=3 \ ,    \\
\noalign{\vspace{1ex}}
\displaystyle{
\frac{ \tau^{2} F_+  F_-  }{  4  }
\coth \! \left( \frac{ \tau F_+ }{2}  \! \right)
\coth \! \left( \frac{ \tau F_- }{2}  \! \right) }
& {\mathrm{for}} &  D=4 \ .
\end{array}
\right.
\label{trdet}
\end{equation}

In order to calculate the two-loop contribution eq.(\ref{2loopEffectivePot}) first we express the gluon propagator eq.(\ref{GluonPropagator}), in terms of the proper time, as
\begin{eqnarray}
\Delta_{\mu\nu}^{ab}(x,y) = \int_0^{\infty}d\tau {e}^{-\tau M_g^2}
\left[\left({e}^{-2i\tau {\cal F}}\right)_{\mu\nu}
\langle x  | {e}^{- \tau \Pi_{\mu}^2} | y \rangle \right]^{ab} \ ,
\quad a,b = 1,2,3 \ ; \label{GluonProperTimeRep}
\end{eqnarray}
where $M_g$ is the gluon mass eq.(\ref{GluonMassCondition}).
(Gluons would be massive in the considering situation, that is, in the confining phase.) The results whose derivation is relegated to Appendix~\ref{Appb} are
\begin{eqnarray}
& & \Delta_{\mu\nu}^{ij}(x,y)
=\left[\left(\cos{\cal C}+{\bbox\epsilon}\sin{\cal C}\right)
\left(\Delta_{\mu\nu}^{1}(x-y)
+{\bbox\epsilon} \ \Delta_{\mu\nu}^{2}(x-y)\right)\right]^{ij} \ ;
\quad  \ i,j=1,2 \ , \label{Delta12} \\
& & \Delta_{\mu\nu}^{33}(x-y)\equiv\Delta_{\mu\nu}^{3}(x-y)
=\delta_{\mu\nu}\int_{0}^{\infty} d{\tau}
{{e}^{-\tau M_g^2}\over (4\pi\tau)^{D/2}}
\exp\left[-{1\over 4\tau }(x-y)^2 \right] \ ,
\label{Delta33}
\end{eqnarray}
and others $=0$. (If we work with an abelian gauge theory, only $\Delta^3_{\mu  \nu }$ term survives. In this sense, we call eq.(\ref{Delta33}) the abelian part.) Here ${\cal C}$ has been given in eq.(\ref{A-no-Definition}), ${\bbox\epsilon}$ is a 2 by 2 antisymmetric matrix $(\epsilon^{12}=-\epsilon^{21}=1)$,
\begin{eqnarray}
\Delta_{\mu\nu}^{1}(x-y)
&=&\int_0^{\infty} d\tau{{e}^{-\tau M_g^2}\over (4\pi\tau)^{D/2}}
(\cos2\tau {\cal F})_{\mu\nu}
\left[{\rm det} \left({\sin\tau {\cal F} \over \tau {\cal F}}\right)_{\mu \nu}
\right]^{-1/2}  \ ,
\nonumber  \\
&&\times
\exp\left[-{1\over 4}(x-y)\cdot\left( {\cal F} \cot\tau {\cal F} \right)\cdot
(x-y)\right]
\label{Delta1}\\
\Delta_{\mu\nu}^{2}(x-y)
&=&-\int_0^{\infty} d\tau{{e}^{-\tau M_g^2}\over (4\pi\tau)^{D/2}}
(\sin2\tau {\cal F})_{\mu\nu}
\left[{\rm det}\left({\sin\tau {\cal F} \over \tau {\cal F}}\right)_{\mu \nu}
\right]^{-1/2}
\nonumber  \\
&&\times
\exp\left[-{1\over 4}(x-y)\cdot\left( {\cal F} \cot\tau {\cal F} \right)\cdot
(x-y)\right] \ .
\label{Delta2}
\end{eqnarray}
Second we need the vacuum polarization tensor eq.(\ref{VacPolTensor}) for 
the two-loop effective potential eq.(\ref{2loopEffectivePot}), whose proper time expression is found as follows:
\begin{eqnarray}
{\rm{Tr}}{\rm{Ln}}\Big[\gamma_{\mu}(\partial_{\mu}
-i{\cal A}_{\mu} T_3 - iQ_{\mu})+m\Big]  
&=&{1\over 2}{\rm{Tr}}{\rm{Ln}}\Big[\gamma_{\mu}\gamma_{\nu}
(\Pi_{\mu}-Q_{\mu})(\Pi_{\nu}-Q_{\nu})+m^{2}\Big]  \nonumber \\
&=&-{1\over 2}\int_{1/\Lambda^2}^{\infty}du u^{-1}e^{-um^2}
{\rm{Tr}}\left(e^{-u(H_{0}+H_{1}+H_{2})}\right)  \ , \label{Q-no-secondOrder1}
\end{eqnarray}
where $H_0$ has been given in eq.(\ref{H_0definition}), and
\begin{eqnarray}
&&H_1 \equiv -2Q_{\mu}\Pi_{\mu}-[\Pi_{\mu},Q_{\mu}]
+i\sigma_{\mu\nu}[\Pi_{\nu},Q_{\mu}]  \ , \\
&&H_2 \equiv Q_{\mu}^2+{i \over 2}\sigma_{\mu\nu}[Q_{\mu},Q_{\nu}] \ , \qquad Q_\mu \equiv Q_\mu^aT_a \ .
\end{eqnarray}
Expand the final expression of eq.(\ref{Q-no-secondOrder1}) with respect to
the quantum field $Q_\mu$ up to the second order to find
\begin{eqnarray}
{\rm{Tr}}\left(e^{-u(H_{0}+H_{1}+H_{2})}\right)  
={\rm{Tr}}\left(e^{-uH_{0}}\right) + I_1 + I_2  \ ,
\end{eqnarray}
with
\begin{eqnarray}
I_1 \equiv -u{\rm{Tr}}(H_{2}e^{-uH_0}) & = & -u{\rm{tr}}\int d^{D}x\langle
x\vert H_{2}e^{-uH_0}\vert x\rangle =  -u{\rm{tr}}\int
d^{D}xH_{2}(x)K(x,x;u) \ ,   \\
H_{2}(x)& \equiv & Q_{\mu}^2(x)-{1 \over 2}\sigma_{\mu\nu}\epsilon^{ab3}
Q_{\mu}^a(x) Q_{\nu}^b(x)T_{3}   \ ,
\end{eqnarray}
and
\begin{eqnarray}
I_2& \equiv &{u\over 2}\int_{0}^{u}du_{1}{\rm{Tr}}
\left(H_{1} e^{-(u_{1}-u)H_{0}}H_{1} e^{-u_{1}H_{0}}\right) = {u^2\over
2}{\rm{Tr}}\int_{0}^{1}dv\left(H_{1} e^{-(1-v)uH_0/2}
H_{1} e^{-(1+v)uH_0/2}\right)  \nonumber \\
&=&{u^2\over 2}\int_{0}^{1}dv\,{\rm{tr}}\int d^{D}xd^{D}y \
\langle x\vert H_{1} e^{-(1-v)uH_0/2}\vert y\rangle\langle y\vert
H_{1} e^{-(1+v)uH_0/2} \vert x\rangle  \nonumber \\
&=&{u^2\over 2}\int_{0}^{1}dv\,{\rm{tr}}\int d^{D}xd^{D}y \
H_{1}(x,y;{{1-v}\over 2}u)K(x,y;{{1-v}\over 2}u)
H_{1}(y,x;{{1+v}\over 2}u)K(y,x;{{1+v}\over 2}u) \ ,
\label{I2representarion} \\
       H_{1}(x,y;\tau) & \equiv &
-iQ_{\mu}(x)\left({{\cal F} \over 2}\cot{\tau {\cal F} \over 2}\right)_{\mu\nu}
\hspace{-2ex}(x-y)_{\nu}
-Q_{\mu}(x){\cal F}_{\mu\nu}(x-y)_{\nu}T_{3}
+i{\cal D}^{ab}_{\mu}Q^{b}_{\mu}(x)T_{a}
+\sigma_{\mu\nu}{\cal D}^{ab}_{\nu}Q^{b}_{\mu}(x)T_{a} \ ,
\end{eqnarray}
where we have utilized the $|x\rangle$- representation such that
\begin{eqnarray}
\langle x\vert \Pi_{\mu}e^{-\tau H_{0}}\vert y\rangle
&=&\left(-i\partial_{\mu}^{x}+{1\over 2}{\cal F}_{\mu\nu}x_{\nu}T_{3} \right)
K(x,y;\tau)
\nonumber  \\
&=&\left({i\over 2}\left({{\cal F} \over 2}\cot{\tau {\cal F}\over
2}\right)_{\mu\nu}\hspace{-2ex}
(x-y)_{\nu}{\bf I}+{1\over 2}{\cal F}_{\mu\nu}(x-y)_{\nu}T_{3}
\right)K(x,y;\tau) \ ,
\end{eqnarray}
with the aid of eq.(\ref{kenel'sExpression}), and made a change of variable from $u_1$ to $v$, $u_1 =
(1+v)u/2$, in the first line of eq.(\ref{I2representarion}). In terms of $I_1, I_2$, the vacuum polarization tensor eq.(\ref{VacPolTensor}) reads
\begin{equation}
\Pi^{ab}_{\mu\nu}
={1 \over 2}\frac{ \delta^2 }{ \delta Q_\mu^a
\delta Q_\nu^b    }\int_{1/\Lambda^2}^{\infty}du u^{-1}e^{-um^2}
\left[I_{1}+I_{2} \right] \ .
\end{equation}
If we write
\begin{eqnarray}
{\Pi}_{\mu\nu}^{ij}(x,y)& = &\left[\left(\cos{\cal C}
+{\bbox \epsilon}\sin{\cal C}\right)
\left(\Pi_{\mu\nu}^{1}(x-y)+{\bbox\epsilon}\Pi_{\mu\nu}^{2}(x-y)\right)
\right]^{ij} \ ; \quad  {\mathrm{for}} \ i,j=1,2 \ ;
\\
& & {\Pi}_{\mu\nu}^{33}(x-y) = \Pi_{\mu\nu}^{3}(x-y) \ , \qquad (\mbox{Abelian Part}) \ ;
\end{eqnarray}
in order to meet the expression of the gluon propagators eqs.(\ref{Delta12}) and (\ref{Delta33}), the two-loop contribution, eq.(\ref{2loopEffectivePot}), is expressed as
\begin{eqnarray}
{\mbox{\Large$v$}}_2^{(D)}&=&  {1\over VT}{e^2\over 2} \int d^{D}x d^{D}y
\Pi_{\mu\nu}^{ab}(x,y)\Delta_{\nu\mu}^{ba}(y,x)   \nonumber \\
&=& e^2 \int {d^D p \over (2\pi)^D}
\left[\Pi_{\mu\nu}^{1}(p)\Delta_{\nu\mu}^{1}(p)-
\Pi_{\mu\nu}^{2}(p)\Delta_{\nu\mu}^{2}(p)
+{1\over 2}\Pi_{\mu\nu}^{3}(p)\Delta_{\nu\mu}^{3}(p) \right] \ .
\label{2LoopEffecPot}
\end{eqnarray}
Again the third term designates the abelian contribution.

The explicit forms of $\Pi_{\mu\nu}^{1} \sim \Pi_{\mu\nu}^{3}$ are found, after performing the Fourier transformation, as follows: in $2+1$ dimensions, put
\begin{eqnarray}
& &  \phi^{(3)}\equiv p \cdot \left({\cos uv{\cal F}/2-\cos u{\cal F}/2 \over u{\cal F}\sin u{\cal
F}/2 }\right)
\cdot  p  \ ; 
\\
& & \alpha_{\mu\nu}^{\pm}\equiv \left({\cos u{\cal F}/2\pm\cos uv{\cal F}/2
\over
\sin u{\cal F}/2}\right)_{\mu\nu}  \ ; \quad
\beta_{\mu\nu}\equiv \left({\sin uv{\cal F}/2 \over \sin u{\cal
F}/2}\right)_{\mu\nu}   \ ,
\end{eqnarray}
and utilize $N_{\mu \nu}$ given in eq.(\ref{Nin3-dim}) with obvious abbreviations such that
\begin{eqnarray}
N^2_{\mu\nu}\equiv N_{\mu\rho}N_{\rho\nu} \ ; \
N^3_{\mu\nu}\equiv N_{\mu\rho}N_{\rho\sigma}N_{\sigma\nu}  \ ; \
(N p)_{\mu}\equiv N_{\mu\nu}p_{\nu} \ ; \ \mbox{etc}.
\end{eqnarray}
to obtain
\begin{eqnarray}
& & \Pi_{\mu\nu}^{1}(p)={ \frac{ F_{+} }{ 2(4\pi)^{3/2} } }
\int_{1/\Lambda^2}^{\infty}du
{\int_0^1}dvu^{1/2}e^{-u (m^2+\phi^{(3)})} \frac{ 1 }{\sinh uF_{+}/2   }  \nonumber
\\  \nonumber
&\times&
\bigg[ -\frac{F_{+}}{\sinh uF_{+}/2} \delta_{\mu\nu}+{F_{+}  \over 4}(\cosh \frac{uF_{+}}{2}    +\cosh \frac{u vF_{+}}{2} ) (N\alpha^-)_{\mu\nu} +{F_{+} \over 2}\cosh \frac{u vF_{+}}{2}  (N\alpha^+)_{\mu\nu}
\\ \nonumber
&& +  \left\{ {F_{+} \over {2\sinh uF_{+}/2}    }(\cosh \frac{uF_{+}}{2}     - \cosh \frac{uvF_{+}}{2}  )^2
+  F_{+}\sinh \frac{uF_{+}}{2}   \right\}(N^2)_{\mu\nu}
-{5F_{+}  \over 4 }(\cosh \frac{uF_{+}}{2}     - \cosh \frac{uvF_{+}}{2}    )(N^{3}\alpha^-)_{\mu\nu}
\\  \nonumber
&& + F_{+} \sinh \frac{uvF_{+}}{2}  (N^2 {\beta})_{\mu\nu}
+ \cosh \frac{uvF_{+}}{2}  ({\beta}p)_{\mu}({\beta}p)_{\nu} - \coth \frac{uF_{+}}{2}  \sinh \frac{uvF_{+}}{2} \delta_{\mu\nu}p\cdot({\beta}p)
\\  \nonumber
&&+ \frac{\cosh uF_{+}/2  +\cosh uvF_{+}/2}{2} \Big\{ \delta_{\mu\nu}p^2-p_{\mu}p_{\nu} + \delta_{\mu\nu}({\alpha^-}p)\cdot(\alpha^-p)-(\alpha^-p)_{\mu} (\alpha^-p)_{\nu} \Big\}
\\  \nonumber
&& +\frac{\cosh uF_{+}/2  - \cosh uvF_{+}/2}{2}\Big\{ \delta_{\mu\nu}(Np)\cdot(Np) -(N^2)_{\mu\nu}p^2+p_{\mu}(N^2p)_{\nu}+(N^2p)_{\mu}p_{\nu}
-3(Np)_{\mu}(Np)_{\nu}
\\  \nonumber
&&+\delta_{\mu\nu}(N{\alpha^-}p)\cdot(N{\alpha^-}p)
-(N^2)_{\mu\nu}({\alpha^-}p)\cdot({\alpha^-}p)
+({\alpha^-}p)_{\mu}(N^2{\alpha^-}p)_{\nu}
+(N^2{\alpha^-}p)_{\mu}({\alpha^-}p)_{\nu}
-3(N{\alpha^-}p)_{\mu}(N{\alpha^-}p)_{\nu} \Big\}
\\  \nonumber
&&+\sinh \frac{uvF_{+}}{2}     \Big\{({\beta}p)_{\mu}(N{\alpha^-}p)_{\nu}+(N{\alpha^-}p)_{\mu}
({\beta}p)_{\nu}  \Big\}
\\
&&+\sinh \frac{uF_{+}}{2}     \Big\{-2\delta_{\mu\nu}p\cdot(N{\alpha^-}p)+p_{\mu}
(N{\alpha^-}p)_{\nu}
+(N{\alpha^-}p)_{\mu}p_{\nu}-(Np)_{\mu}({\alpha^-}p)_{\nu}
-({\alpha^-}p)_{\mu}(Np)_{\nu}  \Big\} \bigg] \ ;
\end{eqnarray}
\begin{eqnarray}
& & \Pi_{\mu\nu}^{2}(p)={\frac{F_{+}}{2(4\pi)^{3/2}}}
{\int_{1/\Lambda^2}^\infty}du {\int_0^1}dv
u^{1/2}e^{-u (m^2+\phi^{(3)})} \frac{1  }{\sinh  uF_{+}/2  }  \nonumber
\\  \nonumber
&\times& \bigg[ {F_{+} \over 4}( \cosh \frac{uF_{+}}{2}  + 3\cosh \frac{uvF_{+}}{2}  )N_{\mu\nu} -{3F_{+}  \over 4}(\cosh \frac{uF_{+}}{2} - \cosh \frac{uvF_{+}}{2}  )(N^3)_{\mu\nu} -{F_{+}}\sinh \frac{uF_{+}}{2}   (N^2 {\alpha^-})_{\mu\nu}
\\  \nonumber
&&+{\cosh uF_{+}/2  +\cosh uvF_{+}/2    \over 2}\Big\{({\alpha^-}p)_{\mu}p_{\nu}-p_{\mu}
({\alpha^-}p)_{\nu} \Big\}
+{\cosh uF_{+}/2 - \cosh uvF_{+}/2    \over 2}\Big\{p_{\mu}(N^2{\alpha^-}p)_{\nu}
\\  \nonumber
&&-(N^2{\alpha^-}p)_{\mu}p_{\nu} +(N^2 p)_{\mu}({\alpha^-}p)_{\mu}-({\alpha^-}p)_{\mu}(N^2 p)_{\nu}
-4N_{\mu\nu}(Np)\cdot({\alpha^-}p)
-(Np)_{\mu}(N{\alpha^-}p)_{\nu}+(N{\alpha^-}p)_{\mu}(Np)_{\nu} \Big\}
\\  \nonumber
&&+\sinh \frac{uvF_{+}}{2}  \Big\{({\beta}p)_{\mu}(Np)_{\nu}-(Np)_{\mu}({\beta}p)_{\nu}
+N_{\mu\nu}p\cdot({\beta}p)  \Big\}
\\
&& -\sinh \frac{uF_{+}}{2}  \Big\{N_{\mu\nu}p^2+N_{\mu\nu}({\alpha^-}p)\cdot({\alpha^-}p) + p_{\mu}(Np)_{\nu}
-(Np)_{\mu}p_{\nu}+({\alpha^-}p)_{\mu}(N{\alpha^-}p)_{\nu}
-(N{\alpha^-}p)_{\mu}({\alpha^-}p)_{\nu} \Big\}\bigg] \ ; 
\label{Pi23d}\\
& & \Pi_{\mu\nu}^{3}(p)={F_{+}\over 2(4\pi)^{3/2}}{\int_{1/\Lambda^2}^\infty}
du{\int_0^1}dv
u^{1/2}e^{-u (m^2+\phi^{(3)})} \frac{1}{\sinh  uF_{+}/2}  \nonumber
\\   \nonumber
&\times&\bigg[{\cosh uF_{+}/2 + \cosh uvF_{+}/2    \over 2} 
(\delta_{\mu\nu}p^2-p_{\mu}p_{\nu})
+\sinh \frac{uF_{+}}{2}    \Big\{(Np)_{\mu}({\alpha^-}p)_{\nu}+({\alpha^-}p)_{\mu}(Np)_{\nu}
\Big\}
\\  \nonumber
&&+{\cosh uF_{+}/2 - \cosh uvF_{+}/2    \over 2} \Big\{p^2(N^2)_{\mu\nu}
-\delta_{\mu\nu}(Np)\cdot(Np)
-p_{\mu}(N^2 p)_{\nu}-(N^2 p)_{\mu}p_{\nu}
+3(Np)_{\mu}(Np)_{\nu} \Big\} \\
&&+\cosh \frac{uF_{+}}{2}    \Big\{({\beta}p)_{\mu}({\beta}p)_{\nu}-{\beta}_{\mu\nu}p\cdot
({\beta}p)
+({\alpha^-}p)_{\mu}({\alpha^-}p)_{\nu} \Big\}   \bigg] \ .
\label{Pi3 3-d}
\end{eqnarray}

Similarly in $3+1$ dimensions, introduce
\begin{eqnarray}
\phi^{(4)}\equiv \frac{\cosh uF_+/2 - \cosh uvF_+/2}{uF_+ \sinh uF_+/2}(I^+ p)&\cdot&(I^+ p)
+\frac{\cosh  uF_-/2 - \cosh  uvF_-/2}{ uF_- \sinh  uF_-/2}(I^- p)\cdot(I^- p) \ ,
\label{Iin4-dim}
\end{eqnarray}
where
\begin{equation}
I^+_{\mu\nu}
\equiv - \frac{ {F_{+}}^2 ({\cal F}^2)_{\mu\nu}-{ F_{-}}^2 (\tilde{\cal
F})^2_{\mu\nu}}
{F_{+}^4-F_{-}^4} \ ;   \qquad  I^-_{\mu\nu}
\equiv  \frac{ {F_{-}}^2 ({\cal F}^2)_{\mu\nu}-{ F_{+}}^2 (\tilde{\cal
F})^2_{\mu\nu}}{F_{+}^4-F_{-}^4} \left( = I^+_{\mu\nu}(F_+ \leftrightarrow F_-) \right) \ ,
\end{equation}
which can be regarded as projection operators, obeying
\begin{equation}
I_{\mu\nu}^{+}+I_{\mu\nu}^{-}=\delta_{\mu\nu} \ ; \quad
(I^\pm)^2_{\mu\nu}=I^\pm_{\mu\nu}=-(N^{\pm})^2_{\mu\nu} \ ; \quad
(I^\pm N^\pm)_{\mu\nu}=N^\pm_{\mu\nu} \ ; \quad
(I^\pm I^\mp)_{\mu\nu}=(I^\pm N^\mp)_{\mu\nu}=0 \ ,
\label{relation of IN}
\end{equation}
where $N^{\pm}$ have been defined by eq.(\ref{Nin4-dim}). Then
\begin{eqnarray}
& & \Pi_{\mu\nu}^{1}(p)={\frac{F_+F_-}{4(4\pi)^2}}{\int_{1/\Lambda^2}^\infty}du
{\int_0^1}dv u e^{-u (m^2+\phi^{(4)})} \frac{1}{\sinh uF_+/2 \sinh uF_-/2} \Big[ \left(\frac{ F_+ \cosh uv F_-/2 }{ \sinh uF_+/2 }-\frac{ F_-\cosh uF_+/2 }{ \sinh uF_-/2}\right)
I^+_{\mu\nu}  \nonumber \\
& & +\left( \cosh \frac{uvF_+}{2}  \cosh \frac{uvF_-}{2} - \coth \frac{uF_+}{2}  \coth \frac{uF_-}{2}  \sinh \frac{uvF_+}{2}  \sinh \frac{uvF_-}{2}  \right) \Big\{(I^{+} p)\cdot(I^{+} p)I^{+}_{\mu\nu}  -(I^{+} p)_{\mu}(I^{-} p)_{\nu} \Big\}  \nonumber \\
& & \hspace{30mm}+ 2\frac{\cosh uF_+/2 \left(\cosh uF_-/2 - \cosh uv F_-/2 \right) }{\sinh^2 uF_-/2}(I^-p)\cdot(I^- p)I^+_{\mu\nu} \nonumber \\
& & \hspace{30mm} -2 \frac{\cosh uv F_-/2 \left(\cosh uF_+/2 - \cosh uv F_+/2
\right)}{\sinh^2 uF_+/2}(I^+ p)_{\mu}(I^+ p)_{\nu}  \nonumber  \\
& & + \left( \sinh \frac{uv F_+}{2} \ \sinh \frac{uv F_-}{2}  - \frac{\left(1-\cosh uF_+/2 \cosh uvF_+/2 \right) \left(1-\cosh u F_-/2 \cosh uv F_-/2  \right)}{\sinh uF_+/2 \sinh u F_-/2} \right) (N^+p)_{\mu}(N^-p)_{\nu} \nonumber  \\
& & \hspace{60mm} + \left( + \leftrightarrow  - \right) \Big] \ ;
\label{Pi1 4-d} \\
& & \Pi_{\mu\nu}^{2}(p) = {\frac{F_+F_-}{4(4\pi)^2}}{\int_{1/\Lambda^2}^\infty}du
{\int_0^1}dv u e^{-u (m^2+\phi^{(4)})} \frac{1}{\sinh uF_+/2 \sinh uF_-/2} \Big[  \frac{F_- \sinh uF_+/2 }{ \sinh uF_-/2 }N^+_{\mu\nu} 
\nonumber \\
& & \hspace{30mm} -2 \frac{ \sinh uF_-/2 \left(\cosh uF_+/2 - \cosh uvF_+/2 \right)}{\sinh^2 uF_+/2}  (I^+ p)\cdot(I^+ p)N^-_{\mu\nu} 
\nonumber \\
& & \hspace{30mm} +  \coth \frac{uF_-}{2}  \sinh \frac{uvF_+}{2} \sinh \frac{uvF_-}{2}
 \Big\{(I^+ p)\cdot(I^+ p)N^+_{\mu\nu} + (I^- p)_{\mu}(N^+ p)_{\nu} 
 -(N^+ p)_{\mu}(I^- p)_{\nu} \Big\}    \nonumber  \\
&& \hspace{30mm} + \frac{ \cosh uvF_-/2 (1-\cosh uF_+/2 \cosh uvF_+/2 )}{\sinh uF_+/2}  \nonumber \\ 
& &  \hspace{30mm} \times \Big\{ (I^- p)_{\mu}(N^+ p)_{\nu}-(N^+ p)_{\mu}(I^- p)_{\nu} 
 -(I^+ p)_{\mu}(N^+ p)_{\nu} +(N^+ p)_{\mu}(I^+ p)_{\nu}  \Big\}
\nonumber  \\
&& \hspace{60mm} + \left( + \leftrightarrow - \right) \Big]  \ ; 
\label{Pi2 4-d} \\
& & \Pi_{\mu\nu}^{3}(p)= {\frac{F_+F_-}{4(4\pi)^2}}{\int_{1/\Lambda^2}^\infty}
du{\int_0^1}dv u e^{-u (m^2+\phi^{(4)})} \frac{1}{\sinh uF_+/2 \sinh uF_-/2} 
\nonumber \\
& & \times  \Big[\left( \cosh \frac{uvF_+}{2}   \cosh \frac{uvF_-}{2} - \coth \frac{uF_+}{2}   \coth \frac{uF_-}{2} 
\sinh \frac{uvF_+}{2}   \sinh \frac{uvF_-}{2} \right) \Big\{ (I^- p)\cdot(I^-
p)I^+_{\mu\nu}-(I^+ p)_{\mu}(I^- p)_{\nu} \Big\} 
\nonumber  \\
 & & \hspace{10mm} +     \frac{2\cosh uF_-/2 \left( \cosh uF_+/2 - \cosh uvF_+/2 \right)}{\sinh^2uF_+/2 }    \Big\{  \!  (I^+ p)\cdot(I^+
p)I^+_{\mu\nu}-(I^+ p)_{\mu}(I^+ p)_{\nu}\Big\} 
\nonumber  \\
& & \left( \frac{(1 - \cosh uF_+/2 \cosh uvF_+/2 )(1 - \cosh uF_-/2 \cosh uvF_-/2 )}{\sinh uF_+/2 \sinh uF_-/2} -\sinh \frac{uvF_+}{2} \sinh \frac{uvF_-}{2} \right) (N^+ p)_{\mu}(N^- p)_{\nu} \nonumber  \\
& &  \hspace{40mm}  + \left( + \leftrightarrow  -  \right) \Big] \ . 
\label{Pi3 4-d}
\end{eqnarray}
(These expressions are so lengthy that we ensure the correctness by checking the gauge invariance of those, that is, the Ward-Takahashi relation in Appendix~\ref{Appd}.)

Armed with these general results, in the following we consider the magnetic background only and proceed to calculate the gap equation in the $2+1$ then $3+1$ dimensions.

\section{The gap equation in $2+1$ dimensions}\label{S3}
\begin{figure}
\centering\leavevmode
\epsfysize=8cm\epsfbox{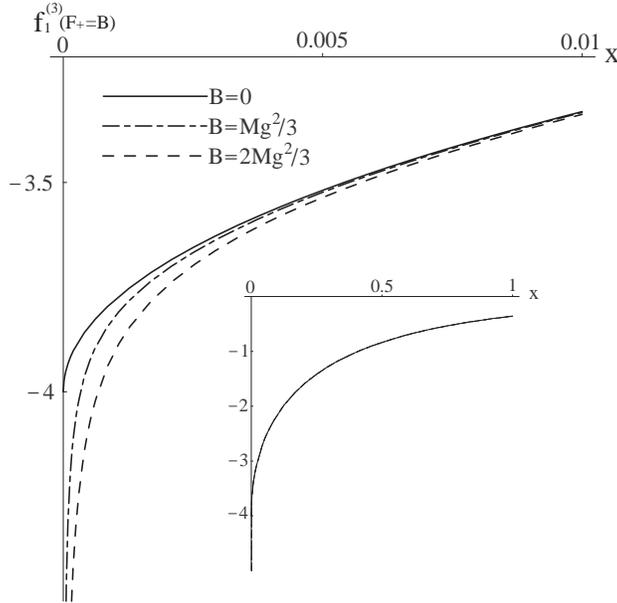}
\caption{One-loop contribution to the gap equation in $2+1$ dimensions. 
Solid line represents $B = 0$, dash-dotted line ${M_g}^2/3$, and dotted lines 
$2{M_g}^2/3$, respectively. In $x > 0.01$, recognized from the small graph, all curves become degenerate. In order to fix the magnitude of the background field, the (dimensionless) gluon mass $y =M_g^2/\Lambda^2$ is set to be $0.01$ .}\label{Fig1}
\end{figure}

When the background is purely magnetic, ${\bbox E}=0$, in view of eq.(\ref{Fs1}), $F_{+}\Longrightarrow B$. A dimensionless quantity,
\begin{equation}
     {\cal B}\equiv{B\over \Lambda^2} \ ,
\end{equation}
is introduced in addition to $x \equiv m^2/\Lambda^2$ in eq.(\ref{TheGap}). (Since the coupling constant $e$ has been included to gauge fields the dimension of gauge fields is always one.) The one-loop contribution to the gap equation eq.(\ref{TheGap}),
\begin{eqnarray}
-\frac{  2(\pi)^{3/2} }{  g^2 \Lambda  }  = f_1^{(3)}(x)  \ ,
\label{3D1-loopGap}
\end{eqnarray}
reads, with the aid of eqs. (\ref{fvRelation}), (\ref{Ir}), and (\ref{trdet}) as,
\begin{eqnarray}
f_1^{(3)}(x)&=&{(4\pi)^{3/2} \over 2\Lambda}
\frac{\partial {\mbox{\Large$v$}}_1^{(3)}}{\partial m^{2}}
=-{\cal B}\int^{\infty}_{1}\!d\tau\,\tau^{-1/2}
e^{-\tau x} \coth \! {\tau {\cal B}\over 2 }  \ ,
\end{eqnarray}
where $\tau$ has been scaled to $\Lambda^2\tau$. We plot $f_1^{(3)}(x)$ in Fig.\ref{Fig1}. It is seen that for a fixed four-fermi coupling $g$, that is, with respect to a (supposed) horizontal line, mass is a monotone increasing function of magnetic field strength. It is also noted that the critical coupling $g_{\rm c}$, defined by $f_1^{(3)}(x=0)$ in eq.(\ref{3D1-loopGap}) for a fixed magnetic field, goes to zero when $B \neq 0$. This phenomenon is so called ``Dimensional Reduction\cite{GMS1}" and is due to the infrared divergence of the effective potential under the background magnetic field\cite{IKT}.

The two-loop contribution is found as, 
\begin{eqnarray}
f_2^{(3)}(x)&=&-\frac{{\cal B}^2 e^2}{2(4\pi)^{3/2}\Lambda}{\int_1^\infty}
\!du{\int_0^{1}}\!dv{\int_0^{\infty}}\!\!\!d\tau\, 
u^{3/2}e^{-ux -\tau y} 
\frac{ \sqrt{{\cal K}} }{  \sinh u{\cal B}/2 }  
\left[\frac{  {\cal M} (q_1\cosh 2\tau {\cal B}+q_2\sinh 2\tau {\cal
B}+q_3) }{ \cosh\tau {\cal B}   }   + {\cal N} q_4  \right] \ ,
\label{3D2LoopGap}
\end{eqnarray}
where $u$ and $\tau$ have been scaled to $\Lambda^2u$ and $\Lambda^2\tau$, 
$y$, defined as 
\begin{eqnarray}
 y \equiv \frac{M_g^2}{\Lambda^2} \ ,
\end{eqnarray} 
is a dimensionless gluon mass,
\begin{eqnarray}
{\cal K}  \equiv  {1 \over u(1-v^2)/4+\tau} & &\ ,  \nonumber  \\ 
{\cal M}\equiv { \sinh u{\cal B}/2 \over \cosh u{\cal B}/2 - \cosh uv{\cal B}/2  + \tanh \tau {\cal B}\sinh u{\cal B}/2} &  , &  \quad {\cal N} \equiv { \sinh u{\cal B}/2 \over \cosh u{\cal B}/2 - \cosh uv{\cal B}/2 + \tau {\cal B}\sinh u{\cal B}/2} \ , \label{abc}
\end{eqnarray}
and
\begin{eqnarray}
&&  q_1  \equiv {{\cal K}\over 2} (\cosh \frac{u{\cal B}}{2}- v \coth \frac{u{\cal B}}{2} \sinh \frac{uv {\cal B}}{2})  \nonumber  \\
& & \hspace{3cm}+ {\cal B} \left( {\cal M} \cosh \frac{u{\cal B}}{2} \cosh \frac{uv{\cal B}}{2}  {\cosh u{\cal B}/2 - \cosh uv{\cal B}/2 \over \sinh^2 u{\cal B}/2 }+  \frac{  (1-\cosh u{\cal B}/2\cosh uv{\cal B}/2) }{\sinh u{\cal B}/2} \right) \ ,  
\label{q1}\\
& & q_2 \equiv  {{\cal K}\over 2}(\sinh \frac{u{\cal B}}{2}  -v\sinh \frac{uv{\cal B}}{2})+ {\cal B}\left( {\cal M} { \cosh uv{\cal B}/2 \over \sinh u{\cal B}/2}(\cosh \frac{u{\cal B}}{2} - \cosh \frac{uv{\cal B}}{2} )
- \cosh \frac{uv{\cal B}}{2}  \right)   \ ,  
\label{q2}\\
& & q_3 \equiv {{\cal K}\over 4}(v^2\cosh \frac{uv{\cal B}}{2} - v \coth \frac{u{\cal B}}{2} \sinh \frac{uv{\cal B}}{2}) + {\cal B}{\cal M}\frac{ \cosh u{\cal B}/2-\cosh uv{\cal B}/2  }{  \sinh^2 u{\cal B}/2  } + {1\over u} \left(\cosh \frac{uv{\cal B}}{2} - {u{\cal B} \over 2 \sinh u{\cal B}/2}\right) \ ,  
\label{q3}\\
& & q_4 \equiv  {{\cal K}\over 4}(\cosh \frac{uv{\cal B}}{2} - v\coth \frac{u{\cal B}}{2} \sinh \frac{uv{\cal B}}{2} )  \nonumber  \\ 
& & \hspace{4.5cm}
+{\cal B}{{\cal N} \over 2}\left( {\cosh u{\cal B}/2-\cosh uv{\cal B}/2 \over \sinh^2 u{\cal B}/2}  +{\cosh uv{\cal B}/2- v\coth u{\cal B}/2 \sinh uv{\cal B}/2 \over 2} \right)   \ .\label{q4}
\end{eqnarray}
$q_1, q_2$, and $q_3$ terms come from the first two terms in eq.(\ref{2LoopEffecPot}), that is, from the nonabelian contribution. Meanwhile $q_4$ represents the abelian contribution. 

\begin{figure}
\centering\leavevmode
\epsfysize=8cm\epsfbox{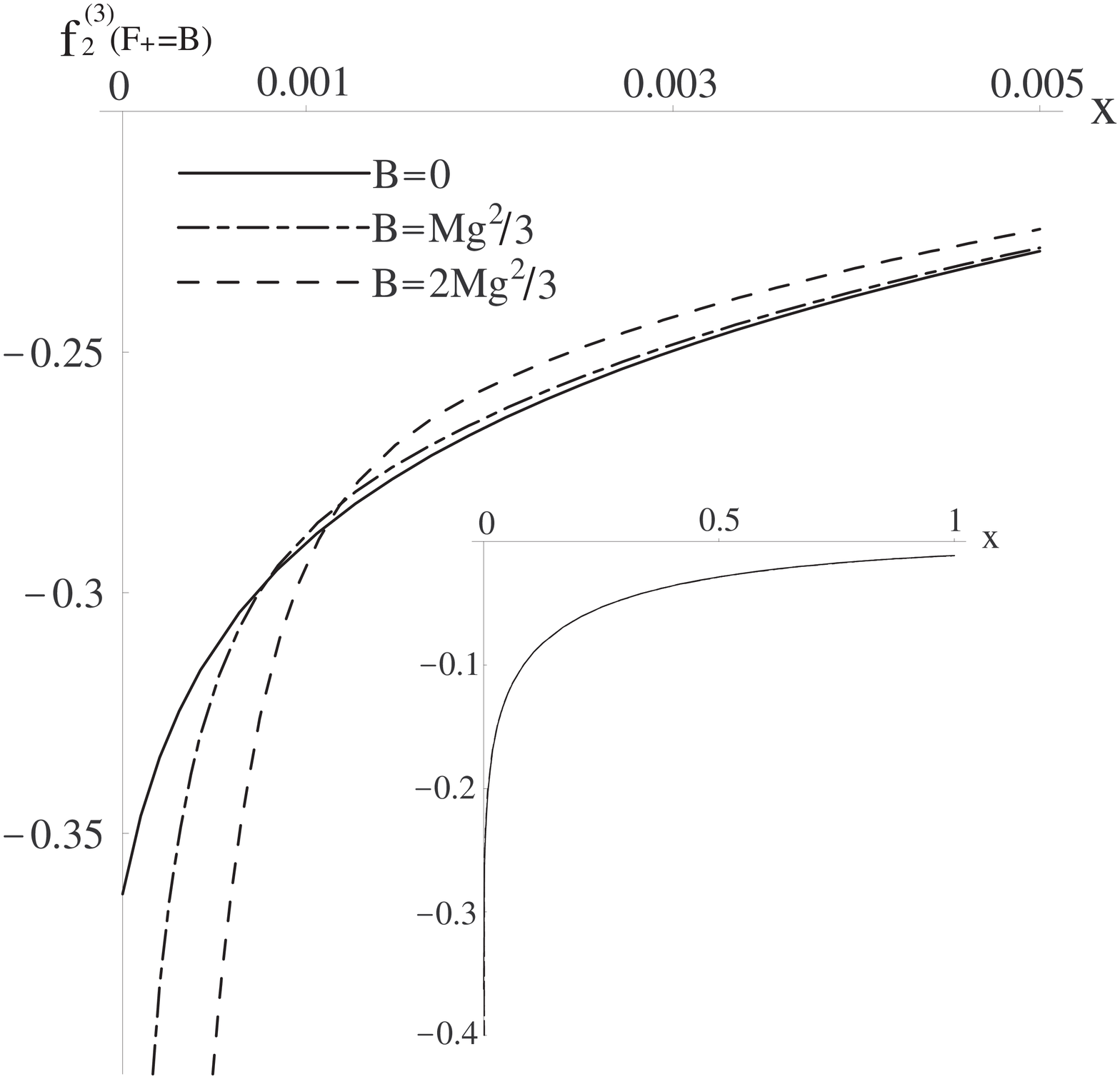}
\caption{Two-loop contribution to the gap equation in $2+1$ dimensions. As in Fig.1, the same line pattern is used. From the smaller graph, we see that all curves become degenerate but remain negative in the whole region $1 \geq x \geq 0$. The graph is drawn by putting $e^2/ (4\pi\Lambda)=0.01 \ , y=0.01$.}\label{Fig2}
\end{figure}

We plot the two-loop part in Fig.\ref{Fig2}. (Our choice of the gauge coupling $e^2/ (4\pi\Lambda)=0.01$ does guarantee the approximation; since $f_2^{(3)}/f_1^{(3)} \sim 0.1$ by comparing the vertical scale between Fig.\ref{Fig1} and Fig.\ref{Fig2}.) From the small graph, we can convince that gluons enhance $\chi$SB as is expected; because all curves remain negative for a whole region, $ 0 \leq x \leq 1$. Note that there is a crossover around $x \sim 0.001$: in the region larger than the crossover, $x$, on some horizontal line (a line
with a fixed four-fermi coupling), is a monotone decreasing function of the 
magnitude of the background field. In the region smaller than that, $x$ is,
however, a increasing function of it, similar to the one-loop case. To see the situation more carefully, we plot the abelian contribution to the gap equation, that is, $q_4$ term in eq.(\ref{3D2LoopGap}).
\begin{figure}
\centering\leavevmode
\epsfysize=8cm\epsfbox{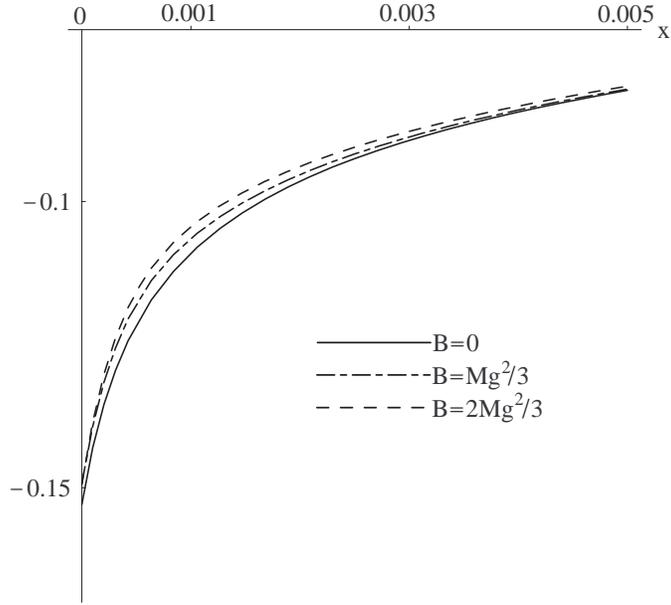}
\caption{Abelian part of the two-loop contribution to the gap equation 
in $2+1$ dimensions. The graph is drawn by putting 
$e^2/ (4\pi\Lambda)=0.01 \ , y=0.01$. }\label{Fig2.1}
\end{figure}
From Fig.\ref{Fig2.1}, $x$, on some horizontal line, is a monotone decreasing function of the background field everywhere for a fixed $g$. Therefore, the increasing tendency of Fig.\ref{Fig2} in $x < 0.001$ comes from the nonabelian parts in eq.(\ref{3D2LoopGap}).

\section{The gap equation in $3+1$ dimensions}\label{S4}

\begin{figure}
\centering\leavevmode
\epsfysize=8cm\epsfbox{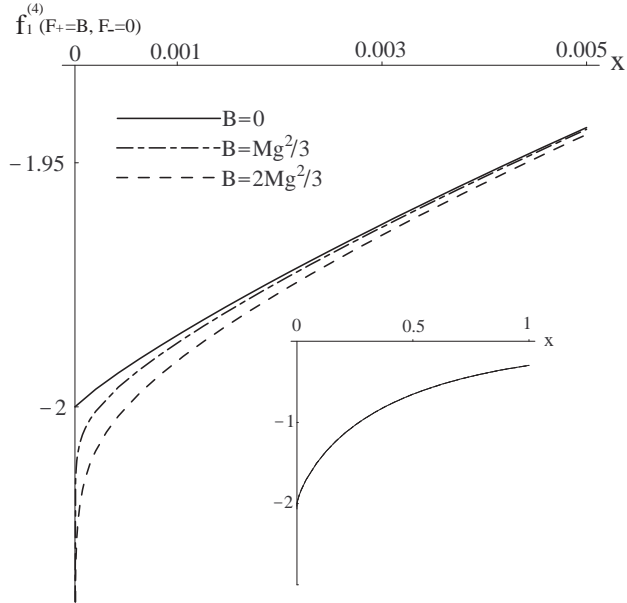}

\vspace{0ex}

\caption{One-loop contribution to the gap equation in $3+1$ dimensions. 
Solid line, dash-dotted line, and dotted line designate $B=0, M_g^2/3$, and
$2M_g^2/3$, respectively. The smaller graph is shown the whole structure, 
$1 \geq x \geq 0$. We again set $M_g^2/\Lambda^2=0.01$.
}\label{Fig3}
\end{figure}

In $3+1$ dimensions, when ${\bbox E}=0$
\begin{equation}
F_{+} \Longrightarrow B   \ ,  \qquad \ F_{-}\Longrightarrow 0  \ .
\end{equation}
Again employing the dimensionless quantity ${\cal B}=B/\Lambda^2$, we have the gap equation of one-loop contribution,
\begin{equation}
-{4(\pi)^{2} \over g^2\Lambda^{2}} = f_1^{(4)}(x)  \ ,  
\end{equation}
with
\begin{eqnarray}
f_1^{(4)}(x)
&=&{(4\pi)^{2} \over 2\Lambda^{2}}
\frac{\partial {\mbox{\Large$v$}}_1^{(4)}}{\partial m^{2}}
=- {\cal B}\int^{\infty}_{1}\!d\tau\,\tau^{-1}
e^{-\tau x}
\coth \! {\tau {\cal B}\over 2 }   \ ,
\end{eqnarray}
which is plotted in Fig.\ref{Fig3}. All curves become degenerate again where $x$ is large. For a fixed four-fermi coupling $g$, mass is a monotone increasing function of magnetic field strength. Moreover the critical coupling goes to zero even under infinitesimal magnetic
fields\cite{GMS1,IKT}.

\begin{figure}
\centering\leavevmode
\epsfysize=8cm\epsfbox{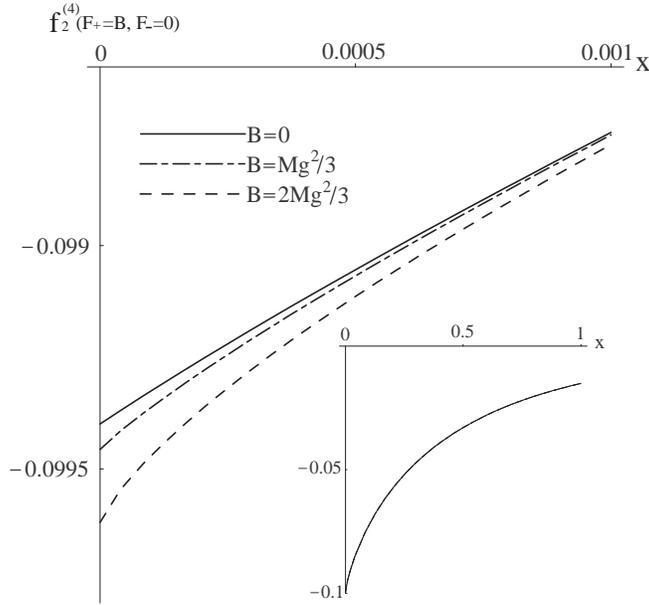}
\caption{Two-loop contribution to the gap equation in $3+1$ dimensions. In the smaller graph the whole structure, $0 \leq x \leq 1$, is shown. The same line pattern is used for different curves by putting ${e^2/ 4\pi}=0.01 $ and $M_g^2/\Lambda^2=0.01$. }\label{Fig4}
\end{figure}

The two-loop contribution is given by
\begin{eqnarray}
f_2^{(4)}(x)&=&-\frac{{\cal B}^2 e^2}{2 (4\pi)^2}{\int_1^\infty}\!du {\int_0^1}
\!dv{\int_0^\infty}\!\!\!d\tau\, 
u\,e^{-ux-\tau y}\frac{  {\cal K} }{  \sinh u{\cal B}/2  }
\left[\frac{  {\cal M} ( p_1\cosh 2\tau {\cal B} +p_2\sinh 2\tau {\cal B} + p_3 )
}{  \cosh \tau {\cal B}  }+ {\cal N}p_4 \right] \ ,
\label{4D2LoopGap}
\end{eqnarray}
where ${\cal K}, {\cal M},$ and ${\cal N}$ are the same as eq.(\ref{abc}) and
\begin{eqnarray}
& & p_1 \equiv  {\cal K}(1-v^2)\cosh \frac{u {\cal B}}{2} + {\cal B}
{\cal M}\left( \cosh \frac{uv{\cal B}}{2} - v\coth \frac{u{\cal B}}{2} \sinh \frac{uv{\cal B}}{2} -{\cosh u{\cal B}/2-\cosh uv{\cal B}/2 \over \sinh^2 u{\cal B}/2}\right) \nonumber  \\ 
& & \hspace{10cm} - {2\over u}\left( \cosh \frac{u{\cal B}}{2} - {u{\cal B}\over 2 \sinh u{\cal B}/2}   \right)  \ ,  \label{p1} 
\\
& & p_2 \equiv {\cal K} (1-v^2)\sinh \frac{u{\cal B}}{2}+ {\cal B} {\cal M}\left({\cosh u{\cal B}/2 \cosh uv {\cal B}/2-1 \over \sinh u{\cal B}/2 }-v\sinh \frac{uv {\cal B}}{2} \right)  -{2 \sinh u{\cal B}/2 \over  u} \ , \label{p2}\\
& & p_3  \equiv {\cal K}\left(\cosh \frac{uv{\cal B}}{2} - v \coth \frac{u{\cal B}}{2} \sinh \frac{uv{\cal B}}{2} -{1-v^2 \over 2}\cosh \frac{uv{\cal B}}{2} \right) + 2{\cal B}{\cal M}\frac{ \cosh u{\cal B}/2-\cosh uv{\cal B}/2}{  \sinh^2 u{\cal B}/2  }  \nonumber  \\ 
& & \hspace{10cm} 
+ {2\over u}\left(\cosh \frac{uv{\cal B}}{2}  - {u{\cal B}  \over 2 \sinh u{\cal B}/2}\right) \ ,  \label{p3}
\\
& & p_4 \equiv {{\cal K}\over 2}\left(\cosh \frac{uv{\cal B}}{2} - v\coth \frac{u{\cal B}}{2} \sinh \frac{uv{\cal B}}{2} +{1-v^2\over 2}\cosh \frac{u{\cal B}}{2} \right)   \nonumber  \\ 
& & \hspace{4cm}  +{\cal B} {{\cal N}\over 2}\left(\cosh \frac{uv{\cal B}}{2} - v\coth \frac{u{\cal B}}{2} \sinh \frac{uv{\cal B}}{2} +{\cosh u{\cal B}/2-\cosh uv{\cal B}/2 \over \sinh^2 u{\cal B}/2}
\right)  \ .\label{p4}
\end{eqnarray}
The graph is shown in Fig.\ref{Fig4}. (The choice of the gauge coupling ${e^2/ 4\pi}=0.01$ again guarantees our approximation; since $f_2^{(4)}/f_1^{(4)} \sim 0.05$. It should be noted that our result of $B=0$ is consistent with that of Kondo, Shuto, and Yamawaki\cite{KSY}.) All curves, shown in the smaller graph, remain negative and become degenerate for $x \geq 0.001$.  
Therefore, gluons enhance $\chi$SB everywhere even in no background $B=0$, which fits our expectation. Contrary to the case in $2+1$ dimensions, mass is a monotone increasing function of magnetic field strength everywhere for a fixed four-fermi coupling $g$.


\section{discussion}\label{S5}

In this paper we discuss the effect of dynamical $SU(2)$ gluons to the gap equation of NJL model under the influence of the constant background magnetic field. The two-loop calculations make expressions considerably complicated but correctness of the results is guaranteed by checking the Ward-Takahashi relation in Appendix~\ref{Appd}. In $3+1$ dimensions, as is seen from Fig.\ref{Fig3} and Fig.\ref{Fig4}, gluons play the same role as fermion in the one-loop, that is, enhance $\chi$SB. Moreover dependence of gluons on the background field is also same as fermion in the one-loop: dynamical mass grows larger as the background
magnetic field becomes stronger. The result is consistent with our expectation
but different from \cite{BGO} where RG with the one-loop calculation was 
employed. In $2+1$ dimensions, the situation is unchanged that gluons enhance 
$\chi$SB even under influence of the background field, contrary to the work of
\cite{GHS} where a different gauge was employed. Dependence of gluons on the background magnetic field, however, is not so simple as in $3+1$ dimensions: as is seen from Fig.\ref{Fig2}, when dynamical mass is tiny the background field increases it, but in a well-broken region, that is, in a region where dynamical 
mass is large, the background field resists a mass to grow. Difference between $2+1$ and $3+1$ dimensions is due to that of the $u$ dependence in eqs.(\ref{3D2LoopGap}) and (\ref{4D2LoopGap}): by making the scale transformation to $u$ and $\tau $ such that 
\begin{eqnarray}
u \mapsto \frac{u}{\cal B}  \ ; \quad  \tau  \mapsto  \frac{\tau }{\cal B}  \  ;  
\end{eqnarray}
quantities, ${\cal K, M , N}$ (\ref{abc}), scale 
\begin{eqnarray}
{\cal K} \mapsto {\cal B K}  \ ;  \quad {\cal M} \mapsto {\cal M} \ ;  \quad  {\cal N}  \mapsto {\cal N} \ ; 
\end{eqnarray}
so that $q_i$ and $p_i \ ;  (i=1 \sim 4)$,  (\ref{q1}) $\sim $ (\ref{q4}) and (\ref{p1}) $\sim $ (\ref{p4}) transform
\begin{eqnarray}
q_i \mapsto {\cal B}q_i  \ ;  \quad
p_i \mapsto {\cal B}p_i  \ . 
\end{eqnarray}
Therefore,
\begin{eqnarray}
f_1^{(3)} \mapsto \sqrt{\cal B}f_1^{(3)}    \ ; \nonumber \\
f_1^{(4)} \mapsto  {\cal B} f_1^{(4)}   \ ; \nonumber  \\ 
f_2^{(4)} \mapsto  {\cal B} f_2^{(4)}   \ ;
\end{eqnarray}
which shows a monotone character with respect to ${\cal B}$, convincing us of the results of Fig.\ref{Fig1} , Fig.\ref{Fig3} , and Fig.\ref{Fig4}, qualitatively. (Because there is still ${\cal B}$ dependence in $f_1^{(3)}, f_1^{(4)}$ , and $ f_2^{(4)}$.) But
\begin{eqnarray}
f_2^{(3)} \mapsto   f_2^{(3)}  \ ,
\end{eqnarray}
which implies that dependence on ${\cal B}$ is due only to the detailed structure of the integrand of the expression (\ref{3D2LoopGap}), that is, we cannot extract a simple monotone behavior in this case. 

The second point we wish to discuss is on instability of gluon functional determinant: we have avoided this by introducing gluon mass $M_g$ which is assumed always bigger than the magnitude of background magnetic field $B$; $M_g^2 >B$. Physically, it is interpreted that the energy of nonabelian particles could become lower and lower as the background magnetic field grows larger and larger. There is no lower limit in the system. The situation is exactly the same in the constant electric field case, where the vacuum becomes unstable due to successive pair productions. We have treated this pathological instability by considering only external electric field whose magnitude $E$ is less than that of dynamical mass squared; $m^2 > E$\cite{IKT}. The point is that the setup itself--``field theories under constant background field''-- is pathological. The system is not closed; energy is continuously supplied from outer environment. However, even in these pathological environments, we could still think about those background  effects to the systems provided that their magnitude is so small.

The final point to discuss is beyond the tree approximation of the auxiliary fields $\sigma$ and $\pi$: in most cases of NJL-study, fermions are assumed to have $N$ components with $N$ being supposed infinite finally. However in the actual situation $N$ is finite so that $O(1/N)$ corrections should be taken into account. A study in a simpler model\cite{kashiwa} says that the approximation becomes more and more accurate if we incorporate higher order terms. Thus going beyond the 1-loop of the auxiliary fields is captivating and the work in this direction is under progress.

\appendix
\section{calculation of kernel by the proper time}\label{Appa}
In this appendix we derive the expression of the kernel (\ref{kernel}),
\begin{equation}
K(x,y,\tau)=\langle x | e^{-\tau H_0} | y \rangle
=e^{\tau\sigma_{\mu\nu}{\cal F}_{\mu\nu}T_3/2 }
\langle x | e^{-\tau \Pi_{\mu}^2} | y \rangle \ . \label{TheKenel}
\end{equation}
Because of the covariantly constant condition (\ref{CovariantConstant}),
the matrix element $\langle x | e^{-\tau \Pi_{\mu}^2} | y \rangle$
can be calculated exactly the same way as the abelian case\cite{FS};
\begin{eqnarray}
\langle x | e^{-\tau \Pi_{\mu}^2} | y \rangle={1\over(4 \pi\tau)^{D/2}}
\exp\left[iT_3{\cal C}\right] \left[{\rm det}
\left({ \sin\tau{\cal F}/2  \over  \tau{\cal F}/2 }\right)_{\mu\nu}
\right]^{-\frac{1}{2}}\exp\left[-\frac{1}{4}(x-y)_\mu\left({{\cal F}\over 2}
\cot{\tau{\cal F}\over 2}\right)_{\mu\nu}\hspace{-2ex}(x-y)_\nu\right] \ ,
\label{matrix element}
\end{eqnarray}
with
\begin{equation}
{\cal C}\equiv-{1\over 2}{\cal F}_{\mu\nu}x_{\mu}y_{\nu} \ .
\label{Definition of C}
\end{equation}
The remaining task is therefore the calculation of ${\displaystyle
\exp\left[{\tau\over 2}\sigma_{\mu\nu}{\cal F}_{\mu\nu}T_3 \right]}$: in
$2+1$ dimensions, the gamma matrices are given as
\begin{equation}
\gamma_{\mu}=\left(\begin{array}{cc}\sigma_{\mu}& 0 \\
0&-\sigma_{\mu}\end{array}\right) \ , \quad \frac{  \sigma_{\mu\nu} }{  2
} \equiv {1\over4i}
[\gamma_{\mu},\gamma_{\nu}]
=\frac{ \epsilon_{\mu\nu\rho}  }{ 2}\left(\begin{array}{cc}\sigma_{\rho}&0\\
0&\sigma_{\rho}\end{array}\right)\equiv\epsilon_{\mu\nu\rho}J_{\rho}\,;\quad
\mu,\nu,\rho=1,2,3  \ ;
\end{equation}
where $J_\mu$'s satisfy,
\begin{equation}
[J_\mu,J_\nu]=i\epsilon_{\mu\nu\rho}J_\rho \  ; \qquad \left\{ J_\mu,J_\nu
\right\} = \frac{  \delta_{\mu \nu} }{ 2 }{\bf I} \ .
\label{J no commutator}
\end{equation}
In terms of $J_\mu$'s,
\begin{equation}
{1\over2}\sigma_{\mu\nu}{\cal F}_{\mu\nu}=E_2J_1
-E_1J_2+BJ_3  \ ; \quad  B \equiv {\cal F}_{12} \ ; {\bbox
E}\equiv ({\cal F}_{13}, {\cal F}_{23})  \ . \label{sigma F 3-d}
\end{equation}
  From eqs. (\ref{J no commutator}) and (\ref{sigma F 3-d}), we obtain
\begin{equation}
\left({1\over2}\sigma_{\mu\nu}{\cal F}_{\mu\nu}T_3 \right)^2
=\left({\sqrt{B^{2}+{\bbox E}^{2}}\over 2}\right)^2{\bf I}
\equiv \left({ F_+\over 2}\right)^2{\bf I} \ .
\end{equation}
Meanwhile,
\begin{equation}
{1\over2}\sigma_{\mu\nu}{\cal F}_{\mu\nu}T_3 = \sigma_{\mu\nu}\frac{  {\cal
F}_{\mu\nu} }{  F_+  } \frac{ F_+  }{ 2 } T_3  \equiv  \sigma_{\mu\nu}
N_{\mu \nu} T_3 \frac{ F_+  }{ 2 }   \ ;  \quad N_{\mu \nu} \equiv \frac{
{\cal F}_{\mu\nu} }{  F_+  } \ . 
\end{equation}
Therefore in $2+1$ dimensions,
\begin{eqnarray}
\exp\left[{{\tau\over 2}\sigma_{\mu\nu}{\cal F}_{\mu\nu}}T_3\right]
=\cosh{\tau F_+\over 2}{\bf I}+\sigma_{\mu\nu}N_{\mu\nu}
\sinh{\tau F_+\over 2}T_3 \ .
\end{eqnarray}

In $3+1$ dimensions, first write
\begin{eqnarray}
& \!  \!  &  \left\{ \begin{array}{l}
\displaystyle{ {F}_{+} \equiv  \frac{ \left\{ \vert{\bbox{B+E}}\vert +
\vert{\bbox{B-E}}\vert\right\}  }{  2  } } \ ;   \qquad {\bbox E} \equiv ({\cal
F}_{14},{\cal F}_{24},{\cal F}_{34})  \\
\noalign{\vspace{1ex} }
\displaystyle{ F_- \equiv \frac{  \left\{\vert{\bbox{B+E}}\vert -
\vert{\bbox{B-E}}\vert\right\} }{  2  } } \ ;  \qquad {\bbox B} \equiv ({\cal
F}_{23},{\cal F}_{31},{\cal F}_{12})
\end{array} \right.   \ ;  \label{Fs2}
\end{eqnarray}
and introduce the antisymmetric tensors,
$N_{\mu\nu}^{\pm}=-N_{\nu\mu}^{\pm}$, such that
\begin{equation}
N_{\mu\nu}^{+} \equiv \frac{ {\cal F}_{\mu \nu} F_+ -
\tilde{\cal F}_{\mu\nu} F_- }{  F_+^2 - F_-^2  } \quad  ;
\quad N_{\mu\nu}^{-}\equiv \frac{\tilde{\cal F}_{\mu \nu}
F_+ - {\cal F}_{\mu \nu} F_- }{  F_+^2 - F_-^2  }  \  :  \quad \tilde{\cal
F}_{\mu \nu} \equiv \frac{ \epsilon_{\mu \nu \lambda \rho}  }{  2  }{\cal
F}_{\lambda \rho} \ , 
\end{equation}
which satisfy
\begin{eqnarray}
\left(N^{\pm}N^{\mp}\right)_{\mu\nu} \equiv N_{\mu\lambda}^{\pm}N_{\lambda
\nu}^{\mp}\!\!& = & \!\! 0 \ ;  \quad \frac{ \epsilon_{\mu \nu \lambda
\rho}  }{ 2 }N_{\lambda \rho}^{\pm} = N_{\mu \nu}^{\mp}.
\label{N'sRelation}  \\
   \left(N^{\pm}_{\mu \nu}\right)^2\!\!& = & \!\! 2 \ ; \label{N'sRelation2}
\end{eqnarray}
where the second relation can be verified by using eqs.(\ref{Fs2}) and
\begin{equation}
\left({\cal F}_{\mu \nu}\right)^2 =\left(\tilde{{\cal F}}_{\mu
\nu}\right)^2 = 2(F_+^2 + F_-^2 ) \ ;  \quad {\cal F}_{\mu \nu}\tilde{{\cal
F}}_{\mu \nu} = 4F_+F_- \ . 
\end{equation}
With the aid of $N_{\mu\nu}^{\pm}$, ${\cal F}_{\mu\nu}$ is expressed as
\begin{equation}
{\cal F}_{\mu\nu}=F_+ N^+_{\mu\nu}+F_- N^-_{\mu\nu} \ ,
\label{F wo N de}
\end{equation}
giving
\begin{equation}
{1\over2}\sigma_{\mu\nu}{\cal F}_{\mu\nu}={1\over2}
(F_+\sigma_{\mu\nu}N^+_{\mu\nu}+F_-\sigma_{\mu\nu}N^-_{\mu\nu}) \ : \quad
\sigma_{\mu\nu}\equiv \frac{  [\gamma_{\mu},\gamma_{\nu}] }{  2i  }  \ .
\end{equation}
By noting
\begin{equation}
[\sigma_{\mu\nu},\sigma_{\lambda\rho}]
=2i(\delta_{\mu\lambda}\sigma_{\nu\rho}-\delta_{\mu\rho}\sigma_{\nu\lambda}
-\delta_{\nu\lambda}\sigma_{\mu\rho}+\delta_{\nu\rho}\sigma_{\mu\lambda}) \ ,
\label{sigma no commutator}
\end{equation}
and eq.(\ref{N'sRelation}), we find
\begin{equation}
[\sigma_{\mu\nu}N^+_{\mu\nu},\sigma_{\lambda\rho}N^-_{\lambda\rho}]=0 \ .
\end{equation}
Therefore
\begin{eqnarray}
\exp\left[{1\over 2}\sigma_{\mu\nu}{\cal F}_{\mu\nu}T_3\right]
=\exp\left[{F_+\over 2}\sigma_{\mu\nu}N^+_{\mu\nu}T_3\right]
\exp\left[{F_-\over 2}\sigma_{\mu\nu}N^-_{\mu\nu}T_3\right] \ .
\end{eqnarray}
Also by noting
\begin{equation}
\{\sigma_{\mu\nu},\sigma_{\lambda\rho}\}
=2(\delta_{\mu\lambda}\delta_{\nu\rho}-\delta_{\mu\rho}\delta_{\nu\lambda}
-\epsilon_{\mu\nu\lambda\rho}\gamma_5) \ ,
\label{sigma no anticommutator}
\end{equation}
and eq.(\ref{N'sRelation2}),
\begin{equation}
(\sigma_{\mu\nu}N_{\mu\nu}^{\pm})^2=4 \ .
\label{sigma to N1}
\end{equation}
Hence
\begin{equation}
\left({F_{\pm}\over 2}\sigma_{\mu\nu}N^{\pm}_{\mu\nu}T_3\right)^2
=\left({F_{\pm}\over 2}\right)^2{\bf I} \ ,
\end{equation}
yielding
\begin{equation}
\exp\left[{F_{\pm}\over 2}\sigma_{\mu\nu}N^{\pm}_{\mu\nu}T_3\right]
=\cosh{F_{\pm}\over 2}{\bf I}+\sigma_{\mu\nu}N_{\mu\nu}^{\pm}
\sinh{F_{\pm}\over 2}T_3 \ .
\end{equation}
Finally utilizing
\begin{equation}
\sigma_{\mu\nu}N_{\mu\nu}^{+}\sigma_{\rho\lambda}N_{\rho\lambda}^{-}
=-4\gamma_5 \ ,
\end{equation}
we obtain
\begin{eqnarray}
\exp\left[{\tau\over 2}\sigma_{\mu\nu}{\cal F}_{\mu\nu}T_3\right]
&=&\left(\cosh\frac{\tau F_+ }{ 2 }\cosh\frac{\tau F_- }{ 2 }
-\gamma_5\sinh\frac{\tau F_+ }{ 2 }\sinh\frac{\tau F_-}{ 2 }\right){\bf I}
\nonumber  \\
&&+\sigma_{\mu\nu}\Big(N_{\mu\nu}^{+}\sinh\frac{\tau F_+ }{ 2 } \cosh
\frac{\tau F_- }{ 2 }+N_{\mu\nu}^{-} \cosh\frac{\tau F_+ }{ 2 }
\sinh\frac{\tau F_- }{ 2 }\Big)T_3  \\
& =&K_0(\tau){\bf I}+K_3(\tau)T_3  \ : \\
K_0(\tau) & \! \equiv  \!  &
\cosh \frac{ \tau F_+ }{ 2 } \cosh \frac{ \tau F_- }{ 2 }
- \gamma_5 \sinh \frac{ \tau F_+ }{ 2 }\sinh \frac{ \tau F_- }{ 2 }  \ ,
\\
K_3(\tau)& \! \equiv \! & \sigma_{\mu \nu}\Big(N_{\mu \nu}^{+}
\sinh\frac{\tau F_+}{ 2 }\cosh\frac{\tau F_-}{ 2 }+N_{\mu \nu}^{-}\cosh
\frac{ \tau F_+ }{ 2 }\sinh \frac{ \tau F_- }{ 2 }\Big)  \ .
\end{eqnarray}

\section{gluon propagator in terms of the proper time}\label{Appb}
In this appendix we show the proper time representation of the gluon propagator
(\ref{GluonPropagator}),
\begin{equation}
(\Delta^{-1})_{\mu\nu}^{ab}=-\delta_{\mu\nu}({\cal D}^2)^{ab}
+2i{\cal F}_{\mu\nu}[\mbox{ad}(T_3)]^{ab}
\ ,
\end{equation}
where we have introduced a slightly different notation from
eq.(\ref{GluonPropagator}),
\begin{equation}
{\cal D}_{\mu}^{ab}=\delta^{ab}\partial_{\mu}-i{\cal A}_{\mu}
[\mbox{ad}(T_3)]^{ab} \ , \quad [\mbox{ad}(T_3)]
\equiv\left(\begin{array}{ccc}0&-i&0\\i&0&0\\0&0&0\end{array}\right) \ .
\end{equation}
The proper time representation is obtained as usual;
\begin{equation}
\Delta^{ab}_{\mu\nu}(x,y)=\int_{0}^{\infty } d{\tau} e^{- \tau M_g^2}
\left[\langle x |e^{-\tau\Pi^2}|y\rangle\right]^{ac}
\left(e^{-2i\tau {\cal F}[\mbox{ad}(T_3)]}\right)_{\mu\nu}^{cb} \ ,
\end{equation}
where
\begin{equation}
\Pi^2 \equiv \left(\Pi_{\mu}^{ab}\right)^2 \ ; \quad
\Pi_{\mu}^{ab}\equiv\delta^{ab}\hat{p}_{\mu}
-{\cal A}_{\mu}(\hat{x})[\mbox{ad}(T_3)]^{ab}
\ ,
\end{equation}
and we have introduced the gluon mass $M_g$ to avoid the tachyonic singularity. For $a=b=3$, it reads
\begin{equation}
\Delta^{33}\equiv\Delta^{3}_{\mu\nu}
=\left(-\delta_{\mu\nu}\partial^2\right)^{-1} \ , \label{GluonPropagator33}
\end{equation}
which is the free propagator. Therefore we obtain eq.(\ref{Delta33}):
\begin{equation}
\Delta_{\mu\nu}^{3}(x-y)
=\delta_{\mu\nu}\int_{0}^{\infty } d{\tau}{  e^{- \tau M_g^2}\over (4\pi\tau)^{D/2}}
\exp\left[-{1\over 4\tau }(x-y)^2 \right] \ .
\end{equation}
In $a,b=1,2$ use $i,j=1,2$ and utilize the result in Appendix~\ref{Appa}  (\ref{matrix element}) and  $[\mbox{ad}(T_3)]^{ij}=-i\epsilon^{ij}$ to find
\begin{eqnarray}
\left[\langle x |e^{-\tau\Pi^2}|y\rangle\right]^{ij}
\hspace{-2ex}={1\over(4 \pi\tau)^{D/2}}(\delta^{ij}\cos{\cal
C}+\epsilon^{ij}\sin{\cal C})
\left[{\rm det}\left({ \sin\tau{\cal F} \over \tau{\cal F} }\right)_{\mu\nu}
\right]^{-\frac{1}{2}}\hspace{-2ex}\exp\left[-\frac{1}{4}(x-y)_\mu
\left({\cal F}\cot\tau{\cal F}\right)_{\mu\nu}(x-y)_\nu\right] \ ,
\label{MatrixElementAdjointRep}
\end{eqnarray}
where ${\cal C}$ is given in eq.(\ref{Definition of C}). Finally by noting that
\begin{equation}
\left(e^{-2i\tau {\cal F}[\mbox{ad}(T_3)]}\right)_{\mu\nu}^{ij}
=\delta^{ij}(\cos 2\tau {\cal F})_{\mu\nu}
-\epsilon^{ij}(\sin 2\tau {\cal F})_{\mu\nu} \ ,
\label{PhaseFactorAdjointRep}
\end{equation}
the relations (\ref{Delta12}) $\sim$  (\ref{Delta2}) are obtained.

\section{Proof that our classical solution satisfies the covariantly
constant condition}\label{Appc}
In this appendix we show that the right-hand side of eq.(\ref{EQUATION})
vanishes:
\begin{equation}
{\rm tr} \Big[\big(\gamma_{\mu}(\partial_{\mu}
-i{\cal A}_{\mu}T_3)+m \big)^{-1}\big(-i\gamma_{\nu}T_a \big)  \Big]
={\rm tr}\Big[ S_{A}(x,x)\big(-i\gamma_{\nu}T_a \big) \Big]=0 \ ,
\end{equation}
where $S_A(x,y)$ is the fermion propagator under the background fields;
\begin{equation}
\big(\gamma_{\mu}(\partial_{\mu}-i{\cal A}_{\mu}T_3)+m \big)S_{A}(x,y)
=\delta^{D}(x-y) \ .
\end{equation}
$S_A(x,y)$ can be expressed, by using proper time method, as
\begin{eqnarray}
S_{A}(x,y)&=&\langle x\vert \big(i\gamma_{\mu}\Pi_{\mu}+m \big)^{-1}
\vert y\rangle
=\langle x\vert(-i\gamma_{\mu}\Pi_{\mu}+m)
(\Pi_{\rho}^2-{1\over 2}\sigma_{\rho\lambda}{\cal F}_{\rho\lambda}T_3
+m^2)^{-1} \vert y\rangle  \nonumber \\
&=&\int_{0}^{\infty} d\tau e^{-\tau m^2}\langle x\vert
(-i\gamma_{\mu}\Pi_{\mu}+m) e^{-\tau H_{0}} \vert y\rangle
\nonumber \\
&=&\int_{0}^{\infty} d\tau e^{-\tau m^2}
\left({1\over 2}\gamma_{\mu}
\left({{\cal F} \over 2}\cot{\tau {\cal F}\over
2}\right)_{\mu\nu}\hspace{-2ex}
(x-y)_{\nu}{\bf I}-{i\over 2}\gamma_{\mu}{\cal F}_{\mu\nu}(x-y)_{\nu}T_{3}
+m \right)K(x,y;\tau) \ ,
\end{eqnarray}
where $K(x,y;\tau) $ is the kernel of eq.(\ref{TheKenel}). Therefore
\begin{eqnarray}
{\rm tr}\Big[ S_{A}(x,x)\big(-i\gamma_{\nu}T_a \big) \Big]
=-2im \ \int_{0}^{\infty} d\tau e^{-\tau m^2}
\frac{ 1 }{ (4 \pi \tau)^{D/2}   }
\left[ {\rm det} \left(\frac{ \sin \tau {\cal F}/2 }{  \tau {\cal F}/2  }
\right)_{\mu \nu}\right]^{-1/2} {\rm tr} \left[ e^{{\tau\over 2}
\sigma_{\mu\nu}{\cal F}_{\mu\nu}T_3} \gamma_{\nu}T_{a} \right] =0 \ ;
\end{eqnarray}
since the trace for the gamma matrices vanishes because the total number of
those is odd.

\section{The Ward--Takahashi relation of vacuum polarization}\label{Appd}
In this appendix it is shown that the vacuum polarization function satisfies
the Ward-Takahashi relation,
\begin{eqnarray}
({\cal D}^{x}_{\mu})^{ab}\Pi_{\mu\nu}^{bc}(x,y)=0 \ ,
\label{WT relation}
\end{eqnarray}
where
\begin{eqnarray}
{\cal D}^{ab}_{\mu}
=\delta^{ab}\partial_{\mu}-\epsilon^{ab3}{\cal A}_{\mu} \ ,
\nonumber
\end{eqnarray}
with ${\cal A}_{\mu}$ being a background field
\begin{eqnarray}
{\cal A}_{\mu}=-{1\over2}{\cal F}_{\mu\nu}x_{\nu} \ .
\end{eqnarray}

The Ward-Takahashi relation (\ref{WT relation}) is separated into
\begin{eqnarray}
\partial_{\mu}\Pi_{\mu\nu}^{3}(x,y)=0 \ ,
\label{WT of Pi3}
\end{eqnarray}
and
\begin{eqnarray}
(\delta^{ij}\partial_{\mu}^{x}-{\epsilon}^{ij3}{\cal A}_{\mu}(x))
{\Pi}^{jk}_{\mu\nu}(x,y)=0 \ ; \quad  {\mathrm{for}} \ i,j,k=1,2 \ .
\label{WT of Pi12}
\end{eqnarray}
The first relation (\ref{WT of Pi3}) can easily be checked by noting eq.(\ref{Pi3 3-d}) and eq.(\ref{Pi3 4-d}) so that the second relation (\ref{WT of Pi12}) must be examined. In view of the fact that ${\Pi}^{ij}_{\mu\nu}(x,y)$ can be written as
\begin{eqnarray}
{\Pi}^{ij}_{\mu\nu}(x,y)=\left[\left(\cos{\cal C}
+{\bbox \epsilon}\sin{\cal C}\right)
\left(\Pi_{\mu\nu}^{1}(x-y)+{\bbox\epsilon}\Pi_{\mu\nu}^{2}(x-y)\right)
\right]^{ij}  \ ; 
\end{eqnarray}
the matrix relation reduces to
\begin{eqnarray}
&&\partial^{x}_{\mu}\Pi_{\mu\nu}^{1}(x-y)
+{\cal A}_{\mu}(x-y)\Pi_{\mu\nu}^{2}(x-y)=0  \ , \label{EvenWT} \\
&&\partial^{x}_{\mu}\Pi_{\mu\nu}^{2}(x-y)
-{\cal A}_{\mu}(x-y)\Pi_{\mu\nu}^{1}(x-y)=0  \ .
\end{eqnarray}
Utilizing the series expansion with respect to the background gauge field, we show, up to $O(F)$, these relations indeed hold: first note that $\Pi_{\mu\nu}^{1}$ and $\Pi_{\mu\nu}^{2}$ are polynomials of even and add powers of $F$ respectively. Thus in $O(1)$ eq.(\ref{EvenWT}) reads
\begin{equation}
\partial_{\mu}\Pi_{\mu\nu}^{1}\Bigr\vert_{F=0}=0 \ ,
\end{equation}
which is fulfilled; since from eq.(\ref{Pi3 3-d}) and eq.(\ref{Pi1 4-d}), $\Pi_{\mu\nu}^{1}$ has been given by
\begin{eqnarray}
\Pi_{\mu\nu}^{1}(p)\Bigr\vert_{F=0}={1\over (4\pi)^{D/2}}
\int_{1/\Lambda^2}^{\infty}du\int_{0}^{1}dv{1-v^2\over
u^{D/2-1}}\exp\left[-u\left(m^2+{1-v^2\over4}p^2\right)\right]
\left\{p^2\delta_{\mu\nu}-p_{\mu}p_{\nu} \right\} \ .
\end{eqnarray}
Next in $O(F)$,
\begin{equation}
\partial_{\mu}\Pi_{\mu\nu}^{2}\Bigr\vert_{O(F)}
-{\cal A}_{\mu}(x-y)\Pi_{\mu\nu}^{1}(x,y)\Bigr\vert_{F=0}=0 \ ,
\end{equation}
which becomes in the momentum space,
\begin{equation}
p_{\mu}\Pi_{\mu\nu}^{2}(p)\Bigr\vert_{O(F)}
+{1\over 2}{\cal F}_{\mu\rho}{\partial \over \partial p_{\rho}}
\Pi_{\mu\nu}^{1}(p)\Bigr\vert_{F=0}=0 \ ,
\label{WT in momentum}
\end{equation}
where $\Pi_{\mu\nu}^2(p)\Bigr\vert_{O(F)}$ can be found from the expression (\ref{Pi23d}) in $2+1$ dimensions. 
\begin{eqnarray}
\Pi_{\mu\nu}^{2}(p)\Bigr\vert_{O(F)}
&=&{1\over (4\pi)^{3/2}}\int_{1/\Lambda^2}^{\infty}du\int_0^1dv
u^{-1/2}\exp\left[-u\left(m^2+{1-v^2\over4}p^2\right)\right]  \nonumber \\
&\times&\bigg[{\cal F}_{\mu\nu}-{u(1-v^2)\over 4}\left(p_{\mu}({\cal F}p)_{\nu}
-({\cal F}p)_{\mu}p_{\nu}+2p^2{\cal F}_{\mu\nu}\right)\bigg]  \ ;\end{eqnarray}
and from eq.(\ref{Pi2 4-d}) in $3+1$ dimensions

\begin{eqnarray}
\Pi_{\mu\nu}^{2}(p)\Bigr\vert_{O(F)}
&=& {1\over (4\pi)^2}\int_{1/\Lambda^2}^{\infty}du\int_0^1dv
\exp\left[-u\left(m^2+{1-v^2\over4}p^2\right)\right]  \nonumber \\
&\times&\bigg[F_+\bigg\{ {1\over u}N_{\mu\nu}^+ - {(1-v^2)\over 2} (I^- p)\cdot(I^- p)N_{\mu\nu}^+   \nonumber \\
&&-{1+v^2\over 4}
\Big( (I^- p)_{\mu}(N^+ p)_{\nu} - (N^+ p)_{\mu}(I^- p)_{\nu}
-(I^+ p)_{\mu}(N^+ p)_{\nu} + (N^+ p)_{\mu}(I^+ p)_{\nu} \Big)  \nonumber \\
&&+{v^2\over 2}
\Big( (I^+ p)\cdot(I^+ p)N_{\mu\nu}^+ + (I^- p)_{\mu}(N^+ p)_{\nu}
-(N^+ p)_{\mu}(I^- p)_{\nu} \Big)\bigg\}
+(+\leftrightarrow-)\bigg] \ , 
\end{eqnarray} 
respectively. With the use of eq.(\ref{relation of IN}), the left-hand side of eq.(\ref{WT in momentum}) is shown to vanish,
\begin{eqnarray}
\mbox{L.H.S}&=&{1\over 2(4\pi)^{D/2}}\int_{1/\Lambda^2}^{\infty}du
\int_{0}^{1}dv
{e^{-um^2}\over u^{D/2-1}}({\cal F}p)_{\nu} \nonumber \\
&\times&{d \over dv}\Bigg[v(1-v^2)\exp\left[-{u(1-v^2)\over 4}p^2\right]
\Bigg] = 0 \ .
\end{eqnarray}
Therefore we can convince ourselves that the Ward-Takahashi relation is satisfied in each order of the background field.



%
%

\pagebreak


%
%

\end{document}